\documentclass[12pt]{iopart}
\usepackage{graphicx}
\usepackage{iopams}  
\usepackage{siunitx}
\usepackage{booktabs}
\usepackage {verbatim}
\usepackage{cite}
\usepackage{todonotes}
\usepackage{nameref}
\usepackage{soul}
\usepackage[justification=justified,labelfont=bf]{caption}

\begin{document}

\title[\small Online asynchronous detection of ErrPs in participants with SCI]{Online asynchronous detection of error-related potentials in participants with a spinal cord injury using a generic classifier}
\author{Catarina Lopes-Dias$^1$,  Andreea I. Sburlea$^1$,  Katharina Breitegger$^2$, Daniela Wyss$^2$, Harald Drescher$^2$,  Renate Wildburger$^2$ and Gernot R. Müller-Putz$^{1,3}$}
\vspace{10pt}
\address{$^1$ \quad Institute of Neural Engineering, Graz University of Technology, Graz, Austria}
\address{$^2$ \quad AUVA-Rehabilitationsklinik Tobelbad, Tobelbad, Austria}
\address{$^3$ \quad BioTechMed, Graz, Austria}
\ead{gernot.mueller@tugraz.at}

\begin{abstract}
A  brain-computer interface (BCI) user awareness of an error is associated with a cortical signature named error-related potential (ErrP). The incorporation of ErrPs’ detection in BCIs can improve BCIs’ performance.

Objective: This work is three-folded. First, we investigate if an ErrP classifier is transferable from able-bodied participants to participants with spinal cord injury (SCI). Second, we test this generic ErrP classifier with SCI and control participants, in an online experiment without offline calibration. Third, we investigate the morphology of ErrPs in both groups of participants.

Approach:  We used previously recorded electroencephalographic (EEG) data from able-bodied participants to train an ErrP classifier. We tested the classifier asynchronously, in an online experiment with 16 new participants: 8 participants with SCI and 8 able-bodied control participants. The experiment had no offline calibration and participants received feedback regarding the ErrPs' detection from its start. {\color{black} For a matter of fluidity of the experiment, the feedback regarding false positive ErrP detections was not presented to the participants but these detections were taken into account in the evaluation of the classifier.} The generic classifier was not trained with the user's brain signals. Still, its performance was optimized during the online experiment with the use of personalized decision thresholds. The classifier's performance was evaluated using trial-based metrics, which consider the asynchronous detection of ErrPs during the entire trials' duration.

Main results: Participants with SCI presented a non-homogenous ErrP morphology, and four of them did not present clear ErrP signals. The generic classifier performed above chance level in participants with clear ErrP signals, independently of the SCI (11 out of 16 participants). Three out of the five participants that obtained chance level results with the generic classifier would have not benefited from the use of a personalized classifier. 

Significance: This work shows the feasibility of transferring an ErrP classifier from able-bodied participants to participants with SCI, for asynchronous detection of ErrPs in an online experiment without offline calibration, which provided immediate feedback to the users.

\end{abstract}

\maketitle

\section{Introduction}

Brain-computer interfaces (BCIs) can assist people with severe motor impairments to operate external devices by converting their modulated brain activity into the control of these devices \cite{Brunner_horizon2020, millan2010,WOLPAW2002}. Although being a promising technology, most BCIs are still error-prone, and the frequent occurrence of errors can spoil the experience of the BCI user. The user's awareness of an unintended response from the device that he/she is controlling is associated with a neural signature known as error-related potential (ErrP)  \cite{schalk2000, ferrez2005}.
 
ErrPs are associated with conflict monitoring and error processing \cite{yeung2004} and can be measured using non-invasive techniques, such electroencephalography (EEG), that are often used for BCIs' control. Therefore, ErrPs can be used to improve BCIs' performance either in a corrective manner, by allowing corrective actions, or in an adaptive manner, by reducing the possibility of future errors  \cite{chavarriaga2014, llera2011use, llera2012adaptive,yousefi2019, artusi}. The real-time detection of ErrPs is pertinent in BCIs used by persons with motor impairments and also in applications targeting healthy users \cite{Zhang_2015,spueler2015error,kreilinger2016single, hakim}. The  incorporation of ErrPs' detection in a BCI promotes a smoother interaction with its user. Nevertheless, this incorporation is not widely investigated.

The use of ErrPs in discrete BCIs, which are controlled in discrete steps, is well established in healthy participants \cite{schalk2000, parra2003,  kreilinger2012error, zander2016neuroadaptive, SalazarGomez2017CorrectingRM,SuKyoung2017,mousavi2017, ehrlich2018, yousefi2019} and  has also been marginally tested in potential end-users of BCIs \cite{Spueler2012}. Still, BCIs are developing in the direction of offering users continuous control of an external device - continuous BCIs \cite{Kobler2018, Ofner2012, Inoue2018, Nooh2011, Allison2012}.  The incorporation of ErrPs in such BCIs requires an asynchronous detection of ErrPs, since the user can realise at any moment, during the control of the device, that an error has occurred. The asynchronous detection of ErrPs has been studied in healthy participants,  both in offline scenarios   \cite{omedes2013,  omedes2014asynchronous, omedes2015analysis, omedes2015asynchronous, spueler2015error,LopesDias2017,Lopes_Dias_2018} and more recently in online scenarios \cite{Lopes-Dias2019}.

A possible explanation for the limited use of ErrPs in BCIs can be linked with most BCIs relying on personalised classifiers, which are constructed with the user's brain signals. Since a considerable amount of data is necessary to reliably train the classifier, personalised classifiers commonly require a long calibration period before the user can receive feedback of its own brain signals. In this manner, combining ErrPs with other controlling signals would imply collecting calibration data for all the different signals, increasing even more the calibration period. Alternatively, using an ErrP classifier that would not require calibration with the user could encourage the integration of ErrPs with other control signals when constructing BCIs. This could be achieved by either transferring an ErrP classifier across different tasks or across different participants. Both options 
have been tested in discrete tasks, in offline conditions 
\cite{iturrate2011,Spueler2012, iturrate2012, Iturrate_2014, SuKyoung2016, SuKyoung2013,Bhattacharyya,ehrlich2018_1,schonleitner2020} and in online conditions \cite{SuKyoung2017}. Recently, the asynchronous detection of ErrPs with a generic classifier has been studied in the context of a continuous task, in offline conditions \cite{LopesDias_generic2019} and in pseudo-online conditions \cite{LopesDias_generic2020}.

Very few works addressed the study of ErrPs in potential BCI end-users and the existing studies are mainly conducted offline. Keyl and colleagues characterized the morphology of ErrPs of spinal cord injured participants and compared it with able-bodied control participants  \cite{keyl}. The ErrP morphology was comparable in the two groups but the ErrPs of the participants in the SCI group showed smaller peak amplitudes. Kumar and colleagues studied ErrPs during post-stroke rehabilitation movements\cite{Kumar2019}. In this work, individual participants did not display very clear ErrP patterns. Sp\"uler and colleagues studied ErrPs in six participants with amyotrophic lateral sclerosis (ALS) in an online experiment,  and showed that the incorporation of ErrPs improves the BCI performance \cite{Spueler2012}. This work also analysed, offline, the transfer of an ErrP classifier from ALS participants to able-bodied control participants.

Our  study has three main aims.  First, we test the feasibility of transferring an ErrP classifier for asynchronous classification from able-bodied participants to potential end-users of BCIs, in particular participants with a high spinal cord injury (SCI). Second, we test the feasibility of using a generic ErrP classifier asynchronously in an online experiment in which both participants with SCI and control participants took part. Third, we investigate the morphology of ErrPs both in participants with SCI and in control participants.

In the work presented here, we recorded EEG from both participants with SCI and control participants while testing asynchronously a generic ErrP classifier in a closed-loop online experiment. The generic classifier has been trained with the EEG data of 15 able-bodied participants from a previous study of ours and was not retrained during the experiment \cite{Lopes-Dias2019}. This allowed us to create an online experiment with no offline calibration period, in which participants received immediate feedback of their brain signals from the very beginning of the experiment onwards.

\section{Methods}

\subsection{Participants}
Sixteen volunteers participated in the experiment, eight of which had a spinal cord injury. The age of the participants with SCI was $37.5 \pm 9.7$ years (mean $\pm$ std). The remaining participants were able-bodied control participants.  Each participant with SCI was matched with a control participant of the same sex and a maximum age difference of 5 years. The control participants were $35.9 \pm 10.8$ years old (mean $\pm$ std). 
 
All participants with SCI had a spinal cord injury between levels C4 and Th2. Table~\ref{table:participants} summarizes the demographical and clinical data of the participants with SCI: age, sex, neurological level of injury (NLI) and  ASIA impairment score (AIS).

\subsubsection{Inclusion and exclusion criteria}

All participants had to be of age between 18 and 65 years. Given that the experimental paradigm required a preserved arm function, all participants with SCI had to have the injury at level C4 or lower. Participants with SCI were excluded if they were artificially ventilated or had major spasms due to possible interference with the EEG measurement. Control participants were required to be able-bodied and with no history of neurological diseases.

\subsection{Ethical approval and measurements}
This study was approved by the local ethics committee
of the Medical University of Graz (ethical approval number 31-501 ex 18/19) and by the Allgemeine Unfallversicherungsanstalt (AUVA) ethical committee. All participants read and signed an informed consent form before the start of the experiment and were paid for their participation. 
The EEG measurements of the participants with SCI took place at AUVA Rehabilitation Clinic Tobelbad and the EEG measurements of the control participants took place at Graz University of Technology.

\begin{table}
	\centering
	\begin{tabular}{ c | c | c | c |c| l}
		\hline
		Participant & Age & Sex & NLI & AIS& Time since injury   \\ \hline
		P1 & 24 & F & C5 & B & $>$ 10 years \\	
		P2   & 29 & M & C7  & C &  $>$ 10 years\\
		P3  & 33 & M & Th2 & D &  $>$ 9 years\\ %
		P4  & 36 & F &  C7 &  B & $>$ 10 years\\
		P5  & 37 & M & C4  &B & $>$10 years \\ 	
		P6   & 39 & M & C6 & B & $>$ 10 years \\ 
		P7   & 48 & M & C4  & B &  6 - 12 months\\	
		P8  & 54 & M &C4 & B & $>$ 1 year \\	\hline	\end{tabular}
	\caption {Summary of the demographical and clinical data of the participants with SCI.}
	\label{table:participants}
\end{table}

\subsection{Hardware and electrodes' layout}

We recorded EEG data with a sampling rate of  \SI{500}{\hertz} using BrainAmp amplifiers and ActiCap caps (Brain Products, Munich, Germany) with
61 active electrodes positioned in a 10-10 layout, as detailed in Figure~1 of the supplementary material. The ground electrode was placed on AFz and the reference electrode was placed on the right mastoid. Additionally, we used 3 EOG electrodes that were placed above the nasion and below the outer canthi of the eyes.

\subsection{Experimental setup}
\begin{figure}[h]
	\centering
	\includegraphics[scale=0.06]{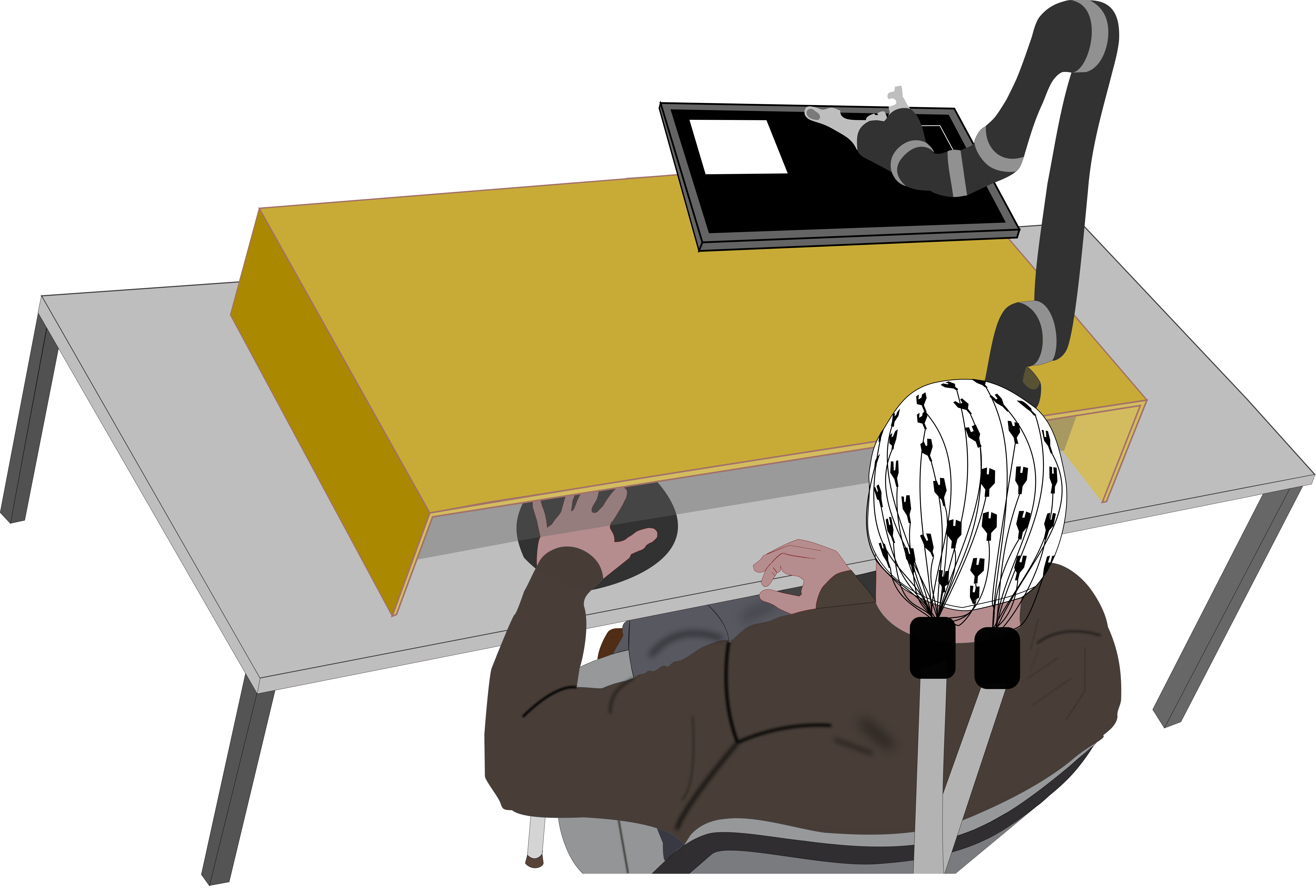}

	 \vspace{0.5cm}
	
	\includegraphics[width=0.6\textwidth]{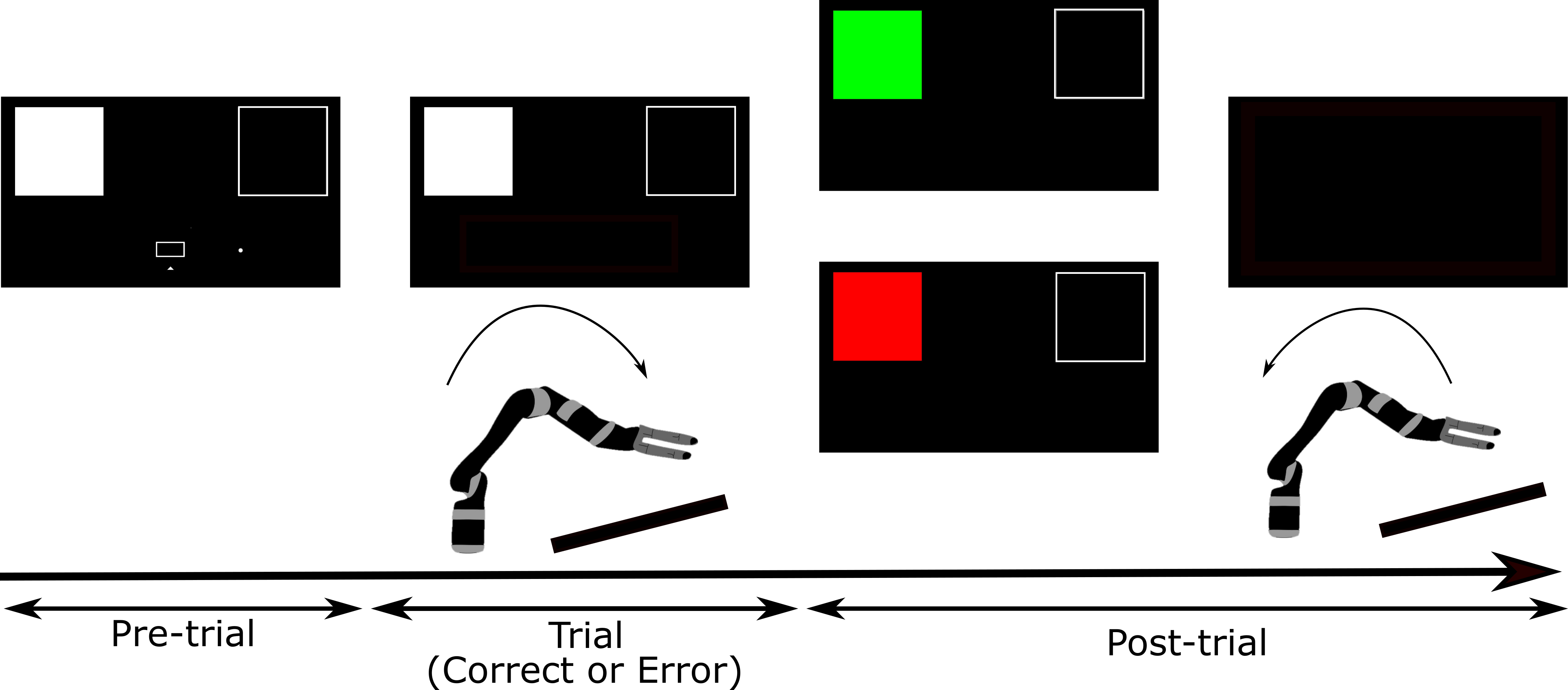}
	\caption{\textbf{Top:} Experimental setup. Participants sat in front of a table, attached which was a robotic arm. The participants controlled the robotic arm during the trials using their hand. \textbf{Bottom:} The experimental protocol displayed on the monitor. During the pre-trial period, the white square represented the target of the coming trial. The small rectangle located centrally, on the bottom part of the screen,  represented the home position of the participant's hand. A trial started when the participants moved their hand to its home position. The participants were instructed to move the robot to the target square during the trials. After the trial (post-trial period), the target changed colour, indicating whether or not it was reached, and the robot automatically returned to its home position.}
	\label{fig:participant}
\end{figure}

Similarly to the experimental setup described in \cite{Lopes-Dias2019}, participants sat in front of a table, on top of which was a wooden 4-sided box, with open sides towards the participant and the tabletop, as depicted in Figure~\ref{fig:participant} (top).
On the ceiling of the box was a Leap Motion device (Leap Motion, San Francisco, United States) that tracked the participants' right or left hand, according to their preferred hand. The position of the Leap Motion on the ceiling of the box was adjusted to the handedness of each participant. The participants kept their hand inside the wooden box. On the right side of the participants, attached to the table, we placed a robotic arm (Jaco Assistive robotic arm - Kinova Robotics, Bonn, Germany). Differently from \cite{Lopes-Dias2019}, a screen monitor was lying on the wooden box, centred in relation to the robotic arm. The monitor was slightly inclined, with a 15-degree angle, to offer the participants a better view  of the screen. This change in relation to \cite{Lopes-Dias2019} was introduced to minimize head and eye movements during the experiment.

\subsection{Controlling the robotic arm}
During the trials, participants could control the robotic arm on a horizontal plane by moving their preferred hand on the tabletop. To reduce the range of the participants' movements, we considered the robot's hand displacement to be three times larger than the participants' hand displacement.   

Many participants with SCI had a very closed fist, due to hand spasticity caused by their injury, and this impaired their hand's recognition by the Leap Motion. When this occurred, we inserted a small object in the participants' hand in order to sustain the hand in a more open position and facilitate its tracking.

\subsection{Experiment overview}
Before the experiment, we recorded one block in which the participant performed eye movements \cite{Kobler2017_eye, Kobler2020}. 
The experiment then consisted of 8 blocks of 30 trials each. \SI{30}{\percent} of the trials of each block were \emph{error trials} (9 trials). The remaining \SI{70}{\percent} of the trials were \emph{correct trials} (21 trials). The sequence of correct and error trials within each block was randomly generated using a uniform distribution. We defined a maximum of 2 consecutive error trials in each block and repeated the randomization procedure until the sequence of trials satisfied this condition. Similarly, the trials of each block were equally split between the right and the left targets. The sequence of targets within each block were randomly assigned using a uniform distribution. We defined a maximum of 3 consecutive trials with the same target in each block and repeated the randomization procedure until the targets' sequence satisfied this requirement.

All the 8 blocks were online blocks: we used a \emph{generic ErrP classifier} in an asynchronous manner to give participants real-time feedback of the \emph{ErrP detections} during the experiment.
For a matter of fluidity of the experiment, we decided not to give participants feedback of false positive ErrP detections, i.e., of the \emph{ErrP detections} that happened when no error had occurred. This decision assured that all participants experienced the same number of errors, which aimed to create a comparable expectation regarding the occurrence of errors across participants. False ErrP detections can occur both in correct and error trials and were considered when evaluating the classifier. The details regarding the generic classifier are described in the section \emph{\nameref{chap:genclass}}.

\subsection{Experimental protocol}
During the pre-trial period, the monitor displayed two squares, on the top part of the screen, each with a 14 cm side. As depicted in Figure~\ref{fig:participant}, one of the squares was filled in white and the other square had no fill. The filled square represented the target of the coming trial. The centres of the squares were 35 cm apart and their midpoint was located 30 cm in front of the home position of the robot's hand.
On the bottom part of the screen was a rectangle, representing the home position of the participant's hand. The position of the participant's hand in relation to its home position was depicted by a dot on the screen.

\bigskip

Participants could decide when to start a new trial and could rest for as long as they needed in between trials. A trial started when the dot entered the rectangle. This ensured that the participant's hand was at a similar position at the beginning of each trial. Participants were instructed to, when they felt ready to start a new trial, position the dot representing their hand below the home position's rectangle, fixate their gaze on the target and finally enter the rectangle from the bottom. This last step ensured a forward movement of the robot. Participants were also asked to keep their gaze fixed at the target during the entire trial in order to prevent eye movements.

The aim of each trial was to move the robot's hand from its home position to the target square. During the trials, only the two squares were displayed on the screen: the white square representing the target and the square with no fill. A trial ended when the robot's hand was above the target or after 6 seconds (time out), in case the target has not been reached. After the end of the trial (post-trial period), the target's colour changed from white to either green or red, for \SI{1.2}{\second}, indicating whether or not the target was reached, respectively. This feedback was always in line with the robot's behaviour. Then, the screen turned black, the robot automatically returned to its home position and a new pre-trial period would start.

\subsubsection{Error trials}
In these trials, the paradigm triggered an error, during the movement of the robot towards the target. The error consisted in interrupting the participant's control of the robot and adding a \SI{5}{\centi\meter} upwards displacement to the robot's hand. The participants perceived the error by noticing the robot stopping and lifting and by realizing that the control of the robot was lost. The errors occurred randomly, when the robot's hand was within 6 to 15 \SI{}{\centi\meter} from its home position, in the forward direction. For every error trial, this distance was drawn from a continuous uniform distribution. In participants with SCI, the error onset occurred, on average, $1.36 \pm 0.14$~\SI{}{\second} after the start of the error trial (mean $\pm$ std). In control participants, the error onset occurred, on average, $1.30 \pm 0.07$~\SI{}{\second} after the start of the error trial  (mean $\pm$ std).

We used the generic ErrP classifier in an asynchronous manner to give participants feedback of the \emph{ErrP detections} occurring after the error onset.  Figure~\ref{fig:metric} illustrates all the possible interactions between the participants and the robot during error trials, taking into account the generic ErrP classifier feedback.
If no ErrP was detected after the error onset, the robot remained still for the rest of the trial. In this situation, the total duration of the trial was 6 seconds and afterwards the target square turned red. Differently, if an ErrP was detected by the classifier after the error onset, the robot's hand lowered \SI{5}{\centi\meter} and the participants regained  its control. The downward movement informed the participants of the ErrP detection and consequent regain of control. Since participants instinctively stopped their hand movement when noticing the error, they were instructed to reinitiate the movement and move the robot's hand to the selected target when regaining control of the robot. To accommodate the extra movement, we added 6 seconds to the maximal trial duration, once the {first ErrP detection after the error onset} occurred. {If the robot reached the target, after the error onset, the target square turned green.}
 Participants did not receive feedback of the false positive detections  occurring during the error trials, i.e., of the \emph{ErrP detections} occurring before the error onset.  Prior to the experiment, participants were informed that errors would occur and were shown the characteristic robot movement associated with error occurrence, i.e., the robot stopping and lifting.

\subsubsection{Correct trials}  
\begin{figure}[h!]
	\centering
	\includegraphics[width=0.9\linewidth]{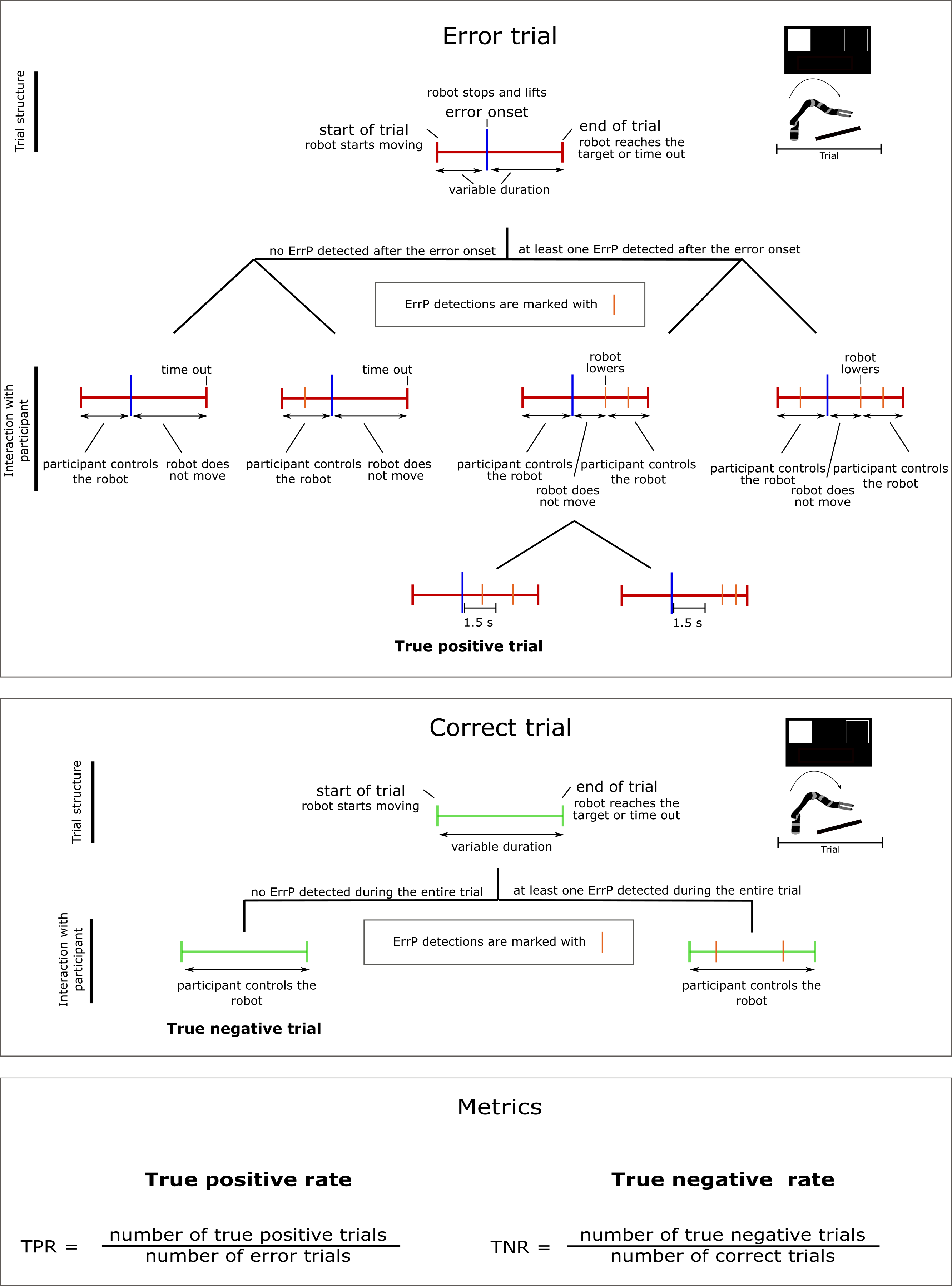}
	\caption{Experimental protocol and metrics. Graphical representation of the trial structure, of the interaction between the participant and the robot during the trials, and of the trial-based metrics used for the evaluation of the classifier. All the occurrences  that are not labelled nor detailed, inherit the corresponding description from the preceding node.}
	\label{fig:metric}
\end{figure}

In these trials, the paradigm did not trigger any error. Participants did not receive feedback of the false positive ErrP detections occurring during the correct trials. Figure~\ref{fig:metric} illustrates all the possible interactions between a participant and the robot during correct trials. Correct trials lasted, on average,
\mbox{ $2.11\; \pm$ \SI{0.17}{\second}} for participants with SCI and $2.05\; \pm $ \SI{0.13}{\second}  for the control participants \mbox{(mean $\pm$ std)}.  All participants reached the target in more than \SI{99.4}{\percent} of the correct trials.

\subsection{Data processing}
Eye movements and blinks were removed online from the EEG data, using the subspace subtraction algorithm \cite{Kobler2017_eye, Kobler2020} and the eye movement data recorded right before the start of the experiment.  
For the online detection of ErrPs with the generic classifier, the EEG data were bandpass filtered between 1 and 10 Hz with a causal Butterworth filter of order 4. For the offline electrophysiological analysis presented here, the EEG data were bandpass filtered between 1 and 10~Hz with a  noncausal Butterworth filter of order 4.

\subsection{Defining events}

In the error trials, we defined the error onset as the moment in which the robot started its upwards displacement once the participant's lost its control. Prior to the experiment, we calculated the robot's delay on 100 uncorrected errors, i.e., the time difference between the error marker and the robot upwards displacement. This resulted in an average delay of $0.225\; \pm$ \SI{0.005}{\second} (mean $\pm$ std). Since the robot's delay was rather stable, we added the average delay to each recorded error marker in order to obtain the error onset.

Correct trials had no clear onset. Therefore, to obtain comparable onsets in correct and error trials for the electrophysiological analysis, we defined a virtual onset for the correct trials  at a time point in which errors could occur in the error trials. For every participant, we defined the virtual onset for his/her correct trials as the average time difference between the error onsets and the start of the corresponding trials. For the participants with SCI, the correct onset occurred, on average, $1.36 \; \pm$ \SI{0.14}{\second}  after the start of the correct trials (mean $\pm$ std). For the control participants, the correct onset occurred, on average,  $1.30 \; \pm$~\SI{0.07}{\second} after the start of the correct trials  (mean $\pm$ std).

\subsection{Generic ErrP classifier}
\label{chap:genclass}
We built a generic error-related potential classifier using the EEG data from 15 able-bodied participants of a previous study for ours \cite{Lopes-Dias2019}. None of these previous participants took part in the experiment described here. The EEG data from those participants were filtered between 1 and 10 Hz using a causal Butterworth filter of order 4. Eye movements were removed from the data using the subspace subtraction algorithm \cite{Kobler2017_eye}. 

For each participant from \cite{Lopes-Dias2019}, we used the 8 calibration runs of the dataset and extracted an epoch with \SI{450}{\milli\second} from every trial. In the error trials, the selected epoch started \SI{300}{\milli\second} after the error onset. In the correct trials, the selected epoch started \SI{300}{\milli\second} after the virtual onset. Hence, our initial features were the amplitudes of the 61 EEG electrodes at all the time points of the \SI{450}{\milli\second} of each epoch.  

In order to remove  outlier epochs, we first  applied principal component analysis (PCA) on the initial features and kept the PCA components that explained \SI{99}{\percent} of the data variability. Then, we removed \SI{1}{\percent} of the correct epochs and \SI{1}{\percent} of the error epochs as outliers. The rejection criterion was based on a large Mahalanobis distance of the rejected epochs within each class type (error or correct) in the PCA space.  After this step, 2475 correct epochs and
	1059 error epochs were kept.

Finally, we repeated the PCA step on the initial feature space, after discarding the outlier epochs, and kept as features the PCA components that preserved \SI{99}{\percent} of the data variability.  This step resulted in 412 PCA components. These components were then used as features to train a shrinkage-LDA classifier with two classes: error and correct \cite{blankertz2011}.  The linear scores of the classifier were transformed into probabilities using a softmax function. The PCA components preserved most of the activity of the original space, as depicted in Figure~2 of the supplementary material. Figure~3 of the supplementary material depicts the classifier pattern, obtained by applying the discriminant feature analysis (DFA) method to the training matrix with 3534 epochs and 412 features \cite{giraldi2008}. The generic classifier remained unchanged during the entire experiment.  In \cite{LopesDias_generic2020},  we showed  that the generic ErrP classifier offers a comparable performance to a personalized ErrP classifier for the asynchronous detection of ErrPs. Therefore, we chose not to retrain the classifier with the participants' own data.

\subsection{ErrP detection}

Similarly to the classifier developed in \cite{Lopes-Dias2019}, the generic classifier developed here was constructed to be used and evaluated in an asynchronous manner.  In the online experiment, the incoming EEG signals were analysed in real-time by the ErrP classifier, which received as input an EEG window of \SI{450}{\milli\second}. Consecutive analyzed windows had a leap of \SI{18}{\milli\second}. The classifier's evaluation of each window resulted in the probability of the analysed window belonging to either class (correct or error). Hence, the classifier produced a probability  output every \SI{18}{\milli\second}, during the entire duration of each block. We defined an \emph{ErrP detection} when two consecutive windows had a probability of belonging to the error class above a certain threshold $\tau$. In \cite{LopesDias_generic2019}, we evaluated offline the asynchronous ErrP detection with the generic classifier and tested the effect of varying the decision threshold. From \cite{LopesDias_generic2019} we concluded that the combination of the generic ErrP classifier with a personalized decision threshold leads to the achievement of better performances. Hence, in this online experiment, we decided to apply this strategy. The procedure to determine the personalized thresholds is described in the section \emph{\nameref{chap:perstau}}.

\subsection{Metrics to evaluate the ErrP classifier}
\label{chapter:metrics}
  To evaluate the performance of the generic classifier, we considered the trial structure of the experiment and the asynchronous nature of the decoding. The proposed metrics assess a trial as successful or unsuccessful, based on the asynchronous detection of ErrPs over the entire trial's duration. This strategy has been applied to the study of asynchronous detection of ErrPs and other event-related potentials, in several other works \cite{omedes2013, Lopez-Larraz2014, omedes2015analysis,omedes2014asynchronous,omedes2015asynchronous, Sburlea_2015,LopesDias2017,pereira2018,Lopes_Dias_2018,LopesDias_generic2019, Lopes-Dias2019,  LopesDias_generic2020}. Figure~2 presents a graphical representation of the metrics proposed here. Correct trials were labelled negative and error trials were labelled positive.
  
\subsubsection{True negative trials}
We defined the true negative trials (TN trials) as the correct trials in which no \emph{ErrP detection} occurred during the entire trial duration. For the classifier's evaluation we considered the true negative rate (TNR): the fraction of correct trials that are TN trials, i.e., that have no \emph{ErrP detections} {\color{black} \footnote{ {\color{black} The metrics TNR and TPR used here address the asynchronous detection of ErrPs in a trial-based scenario and are not directly comparable with the TPR and TPR definitions commonly used in time-locked classification.}}}.
   
\subsubsection{True positive trials}
We defined the true positive trials (TP trials) as the error trials in which no ErrP detection occurred before the error onset and at least one ErrP detection occurred within the \SI{1.5}{\second} after the error onset. For the classifier's evaluation we considered the true positive rate (TPR): the fraction of error trials that were TP trials. {An additional metric, \emph{ErrP detection rate} (EDR), considering only the ErrP detections within the 1.5 s after the error onset, is defined in Figure~5 of the supplementary material, where its relation with the TPR is detailed {\color{black} \ddag}.

\subsubsection{Chance level} 
To calculate the chance level for TNR and TPR we performed several  classifications with a classifier in which the training labels were randomly permuted (500 times for the evaluation of the online detection with the generic classifier and 50 times for the evaluation of the offline cross-validation with a personalized classifier). Furthermore, we used permutation based $p$-values to present the significance of the classification results obtained with the generic ErrP classifier \cite{good2000,ojala2010}.

{\color{black}
\subsection{False activation rate}
The false activation rate (FAR) is the percentage  of  1-second long intervals that  are contaminated  with  at  least  one  false positive ErrP detection \cite{fatourechi}. For this evaluation, we considered the entire duration of correct trials and the period before the error onset in error trials. These periods were divided into 1-second long intervals and these intervals were evaluated for the presence of false positive ErrP detections.
}
\subsection{Tailoring  the decision threshold of the generic classifier to each participant}
\label{chap:perstau}
In \cite{LopesDias_generic2019} we evaluated offline the asynchronous detection of ErrPs with a generic classifier similar to the one described here. There, we observed that the decision threshold $\tau$ that maximized the group performance was $\tau = 0.7$. Moreover, we also concluded that in order to optimize the individual performance with the generic classifier, participants benefit from the use of a personalized threshold. 
 Therefore, in this experiment, we decided to initiate the generic classifier with $\tau = 0.7$ in the first block. This enabled us to skip the offline calibration and allowed us to give participants immediate feedback of their ErrP detections. Afterwards, we tailored $\tau$ to each participant. After each of the first 3 blocks, we performed offline an asynchronous classification with the generic ErrP classifier on all the available data and tested thresholds between 0 and 1 in steps of 0.025. For each of the 41 thresholds analysed, we calculated the corresponding TPR and TNR. The TNR and TPR curves were further smoothed using a moving average with 7 samples. The smoothed curves were named smooth TPR and smooth TNR. For every participant, we chose the threshold that maximized the product of the smooth TPR and the smooth TNR. This was considered the threshold that maximized performance and it was used in the next block. From block 4 onwards, the generic ErrP classifier was combined with the threshold $\tau$ obtained after the third block. The generic ErrP classifier was not retrained with the participants' data and only the decision threshold was updated based on the data.

\subsection{Evaluation of the generic ErrP classifier}
  
We stopped tailoring $\tau$ to each participant after the third block because we wanted to collect a substantial amount of data in unchanged conditions. From blocks 4 \mbox{to 8}, all participants used the generic classifier with a fixed but personalized threshold. Therefore, we only use the data from blocks 4 to 8 to evaluate the performance of the generic classifier, ensuring comparable conditions across the participants.

\subsection{Personalized ErrP classifier}
  
In order to evaluate, offline, the performance of a personalized classifier,  we performed 10 times a 5-fold cross-validation in the entire dataset of each participant, where a personalized classifier was tested in an asynchronous manner in each fold. There, we also tested all thresholds from 0 to 1 in steps of 0.025.  For every participant, we obtained, in each fold, a TPR  and a TNR for every threshold tested. For every participant, we averaged the TPR and TNR of the 50 iterations in the cross-validation, obtaining an average TPR and an average TNR per participant. Finally,  we selected the threshold that maximized the product of the average TPR and the average TNR, for every participant. The evaluation of the personalized classifier followed the metrics defined in the section \emph{\nameref{chapter:metrics}}.

\section{Results}

\subsection{Neurophysiology}

The electrophysiological results presented here comprise the entire recorded dataset.
Figure~\ref{fig:ga} shows the grand average correct and error signals at channel FCz (green and red lines, respectively) for participants with SCI and control participants. The green and red shaded areas depict the \SI{95}{\percent} confidence interval for the grand average signals. The vertical line at $t=$ \SI{0}{\second} depicts the error onset of the error trials and the virtual onset of the correct trials. For the participants with SCI, the grand average error signal displays a negativity, with peak amplitude of \SI{-2.4}{\micro\volt} at time $t=$ \SI{0.154}{\second} after the error onset, followed by a positivity, with  peak amplitude of \SI{3.8}{\micro\volt} at time $t=$ \SI{0.332}{\second}. For the control participants, the grand average error signal displays a negativity, with peak amplitude of \SI{-5.5}{\micro\volt} at time $t=$ \SI{0.176}{\second} after the error onset, followed by a positivity, with peak amplitude of \SI{5.8}{\micro\volt} at time $t=$ \SI{0.334}{\second}.
 The grand average correct signal displays no particular peaks, both in participants with SCI and control participants. Figure~\ref{fig:ga} displays also the topographic plots of the grand average correct and error signals at the time points of the peaks of the grand average error signal.  
 
 \begin{figure}[h]
 	\centering
 	\includegraphics[width=1.0\textwidth]{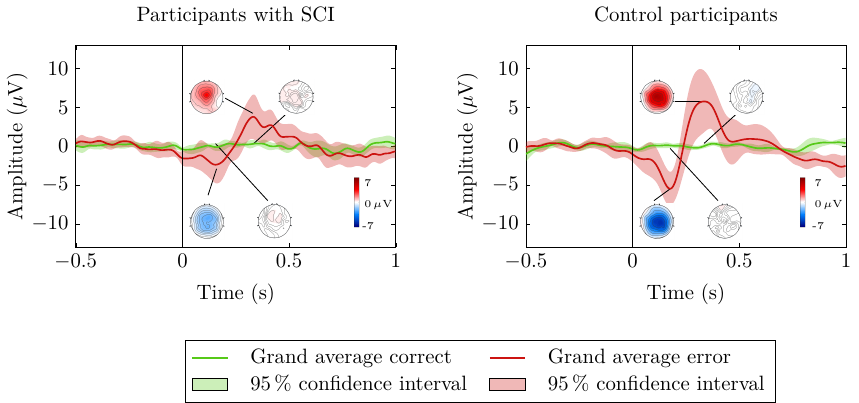}
 	\caption{Grand average correct and error signals at channel FCz (green and red solid lines, respectively) for participants with SCI and control participants. The shaded areas represent the \SI{95}{\percent} confidence interval of the grand average curves. The vertical black line at $t=$ \SI{0}{\second} represents the error onset of the error trials and the virtual onset of the correct trials. The figure displays also the topographic plots of the grand-average correct and error signals at the time points of the peaks in the grand average error signal.}
 	\label{fig:ga}
 \end{figure}

\begin{figure}[h!]
	\centering
	\includegraphics[]{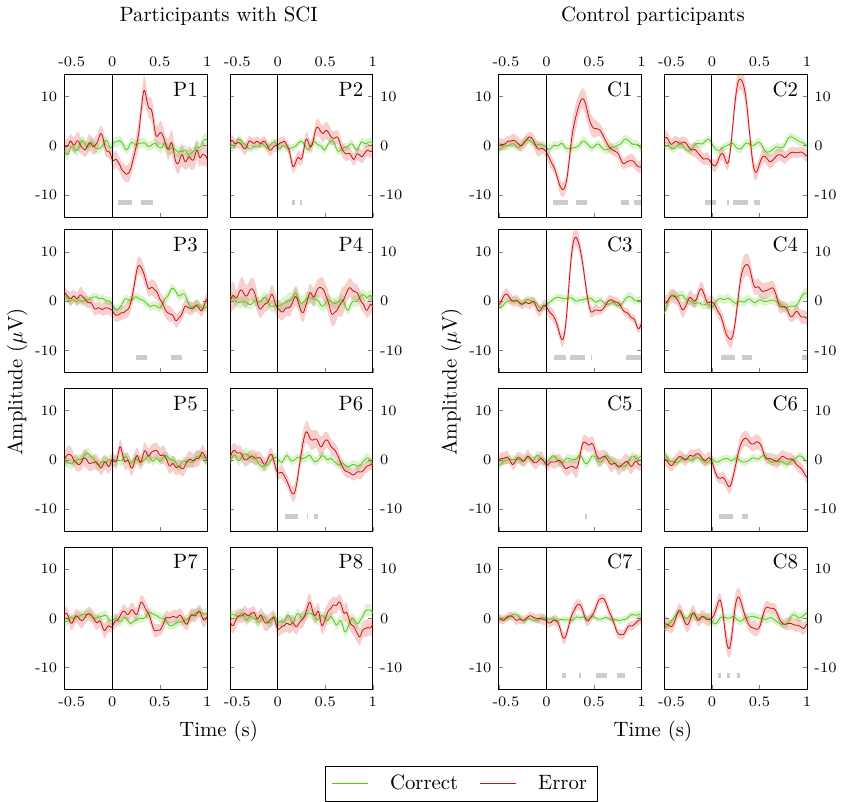}
	\caption{Average correct and error signals at channel FCz  of every participant (green and red lines, respectively). The shaded areas represent the \SI{95}{\percent} confidence interval of the average signals. The  black line at $t=$ \SI{0}{\second} represents the error onset of the error trials and the virtual onset of the correct trials. The grey regions indicate the time points in which correct and error signals were statistically different (Wilcoxon ranksum tests, Bonferroni corrected, with $\alpha=0.01$). }
	\label{fig:errp_ss1}
\end{figure}

As the morphology of the error signals was not homogeneous across participants, we found it relevant to also present the electrophysiological results of the individual participants. Figure~\ref{fig:errp_ss1} displays the average correct and error signals at channel FCz (green and red lines, respectively) of every participant. The green and red shaded areas depict the \SI{95}{\percent} confidence interval for the average signals. The grey areas indicate the time points in which correct and error signals were statistically different (Wilcoxon ranksum tests, Bonferroni corrected, with $\alpha=0.01$). Figures~6 and 7 of the supplementary material depict the topographic plots of the average correct and error signals of every participant at different time points.

\subsection{Adaptation of the classifier's threshold in the first three experimental blocks}

   \begin{figure}[h!]
   	\centering
   	\includegraphics[width=0.9\textwidth]{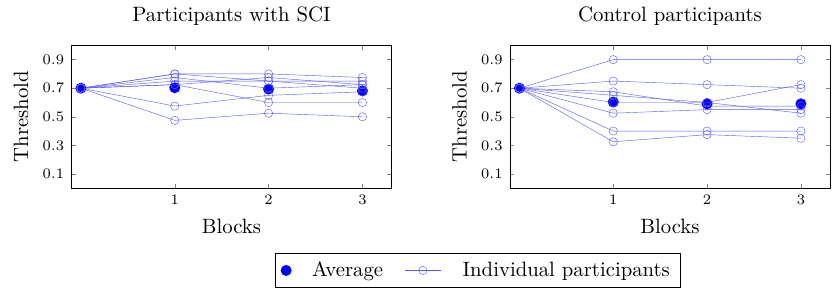}
   	\centering
   	
   	\vspace{0.5cm}
   	
   	\includegraphics[width=0.8\linewidth]{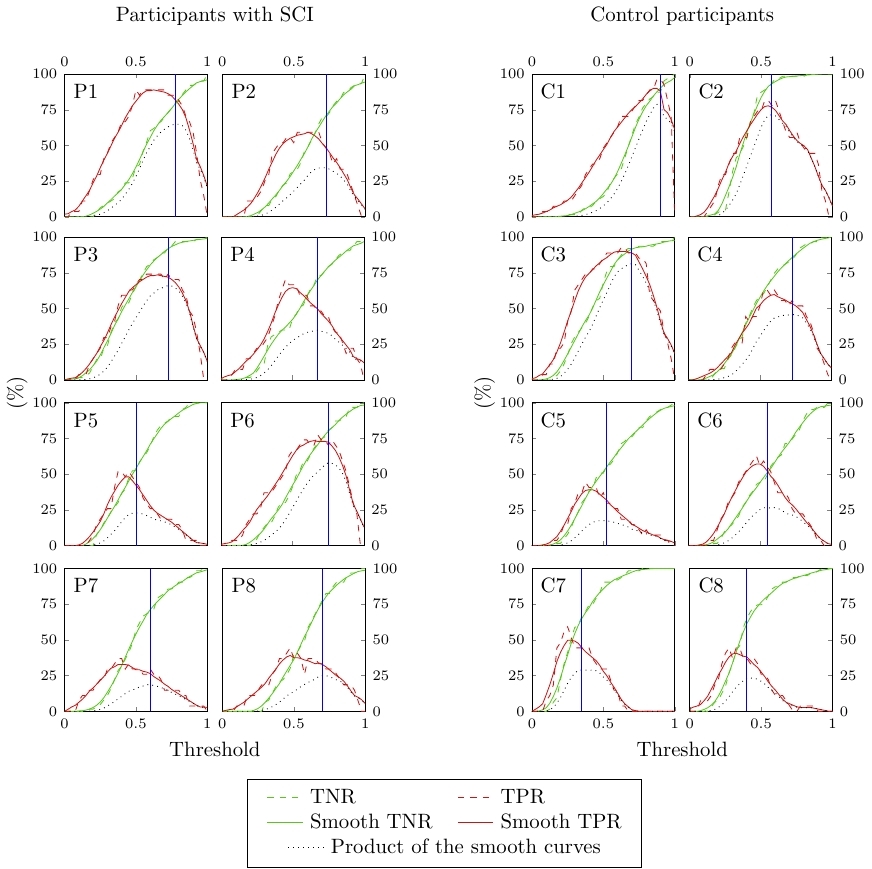}
   	\caption{ Optimization of the decision threshold used with the generic ErrP classifier. \textbf{Top:} Evolution of the decision threshold: Initial threshold  ($\tau=0.7$) and the calculated thresholds after each of the first 3 blocks, for every participant. \textbf{Bottom:} TNR and TPR obtained offline, after the third block (dashed green and red lines, respectively) and the corresponding smooth curves (green and red solid lines). The blue line represents the threshold that maximizes the product of the smooth curves, which is represented with a black dotted line.}
   	\label{fig:threshold}
   \end{figure}

This experiment required no offline calibration and the participants received feedback regarding their ErrP detections from its very beginning. This was possible by combining the generic ErrP classifier with a generic decision threshold ($\tau = 0.7$) for the first experimental block. Still, we used the first three experimental blocks to reach a fixed personalized decision threshold. After each of the first three blocks, we updated the decision threshold $\tau$ in order to maximize the participant's performance.  Hence, participants used a generic classifier combined with a personalized decision threshold from block 2 onwards. Figure~\ref{fig:threshold} (top) depicts the initial threshold ($\tau=0.7$) and the calculated thresholds after each of the first 3 blocks, for every participant. At the end of block 3, the average threshold was $\tau=0.68$ for the participants with SCI and $\tau=0.59$ for the control participants. Figure~\ref{fig:threshold} (bottom) shows the TNR and TPR obtained offline after block 3, for all the tested  thresholds (green and red dashed lines, respectively). It also shows the smooth TNR and smooth TPR, obtained with a moving average (green and red solid lines). The black dotted line depicts the product of these smooth curves and the blue vertical line indicates the threshold that maximizes it. This is the decision threshold used for every participant from blocks 4 to 8.

\subsection{Evaluation of the online asynchronous classification with a generic ErrP classifier}

 \begin{figure}[h]
 	\centering
 	\includegraphics[width=1.0\textwidth]{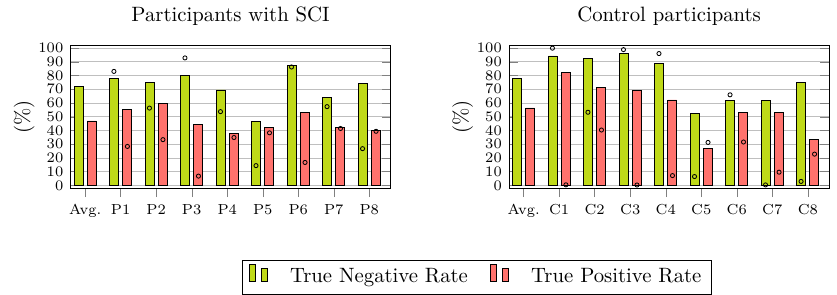}
 	
 	\vspace{0.5cm}
 	
 	\includegraphics[width=0.8\linewidth]{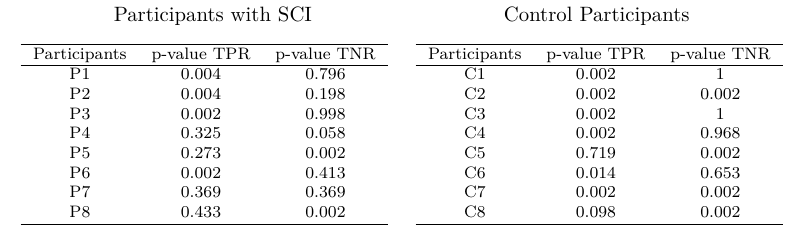}
 	\caption{Evaluation of the generic ErrP classifier. \textbf{Top:} Classification results in terms of true positive rate (TPR) and true negative rate (TNR). The circles on the individual bars represent the chance-level of the corresponding metrics.
 		\textbf{Bottom:} Permutation based $p$-value regarding the significance of the results obtained with the classifier.}
 	\label{fig:genclass}
 \end{figure}

To evaluate the asynchronous classification results obtained with the generic ErrP classifier during the experiment, we only considered the data of the last five blocks of the experiment, i.e., from blocks 4 to 8, since no parameters were changed during these blocks.

 Figure~\ref{fig:genclass} (top) depicts the classification results obtained with the generic classifier in terms of true positive rate (TPR) and true negative rate (TNR). For participants with SCI, we obtained an average TPR of \SI{46.9}{\percent} and an average TNR of \SI{71.9}{\percent}. For control participants, we obtained an average TPR of \SI{56.4}{\percent} and an average TNR of \SI{77.9}{\percent}. The circles on the individual bars represent the chance level of the corresponding metrics. The chance level results for each participant were obtained by averaging the classification results of 500 classifiers in which the training labels were randomly permuted and by considering the final participant-specific threshold, as depicted in  Figure~4 of the supplementary material. Figure~\ref{fig:genclass} (bottom) presents the permutation based $p$-values regarding the significance of the classification results \cite{good2000,ojala2010}. Figure~5 of the supplementary material depicts the comparison between the metrics TPR and EDR. {\color{black} Table~1 of the supplementary material presents the  false activation rate (FAR) in correct and error trials.}

{\color{black} Figure~\ref{fig:asynchronousexample} illustrates the online asynchronous detection of ErrPs and the trials' offline evaluation for participant C1. The dark grey areas represent the trials and the white marks within them represent the ErrP detections. The narrow rectangles  colour code the trials' offline evaluation. In these rectangles, trials successfully classified (true positive trials and true negative trials) are coded in white and trials with false positive ErrP detections are coded in grey. The error trials with no ErrP detection are coded in black.}

 \begin{figure}[h!]
	\centering
	\includegraphics[width=0.95\linewidth]{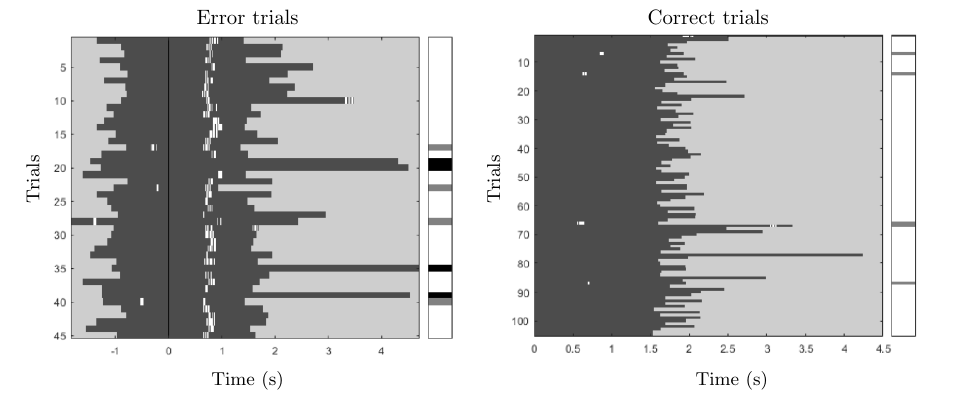}
	\caption{{\color{black}Online detection of ErrPs and trials' offline evaluation for participant C1. Left: Error trials, aligned to the error onset (black vertical line). Right: Correct trials, aligned to the start of the trial. The dark grey areas represent the trials and the white marks within them represent the ErrP detections. The narrow rectangles  colour code the trials' offline evaluation. In these rectangles, trials successfully classified (true positive trials and true negative trials) are coded in white and trials with false positive ErrP detections are coded in light grey. Error trials with no ErrP detections are coded in black. 
		}}
	\label{fig:asynchronousexample}
\end{figure}

\subsection{Offline evaluation of the asynchronous ErrP classification with a personalized classifier}

To evaluate offline the asynchronous classification results with a personalized classifier, we considered the 8 experimental blocks and performed 10 times a 5-fold cross-validation. As this evaluation was done offline, we tested thresholds from 0 to 1 with a leap of 0.025 and the results obtained are shown in function of the threshold $\tau$.

 Figure~\ref{fig:cv} depicts the grand average TNR and TPR (green and red solid lines, respectively) as well as the grand average chance level for TNR and TPR (green and red dashed lines, respectively) in function of the threshold. The shaded areas represent the \SI{95}{\percent} confidence intervals of the grand average curves. The chance level curves were obtained by performing 10 times a 5-fold cross-validation with 50 classifiers in which the labels of the training trials were randomly permuted.
 
  \begin{figure}[h]
 	\centering
 	\includegraphics[width=1.0\textwidth]{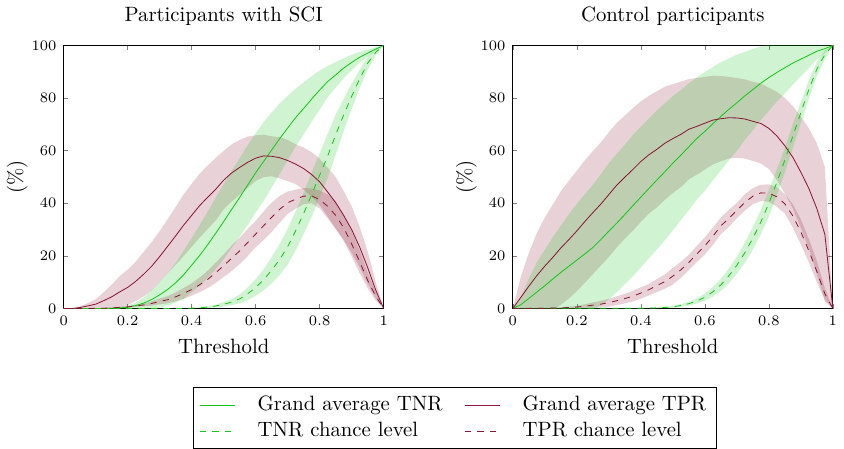}
 	\caption{Evaluation of the personalized ErrP classifier. Grand average TNR and TPR (green and red solid lines, respectively) and grand average chance level TNR and TPR (green and red dashed lines, respectively) in function of the threshold. The shaded areas indicate the \SI{95}{\percent} confidence interval for the grand average curves.}
 	\label{fig:cv}
 \end{figure}

 Figure~\ref{fig:cv_ss} depicts, for every participant, the average TNR and TPR (green and red solid lines, respectively) and the chance level TNR and TPR (green and red dashed lines, respectively). The blue vertical line indicates the threshold that maximizes the individual performance with the personalized ErrP classifier.

\begin{figure}[h]
	\centering
		\includegraphics[]{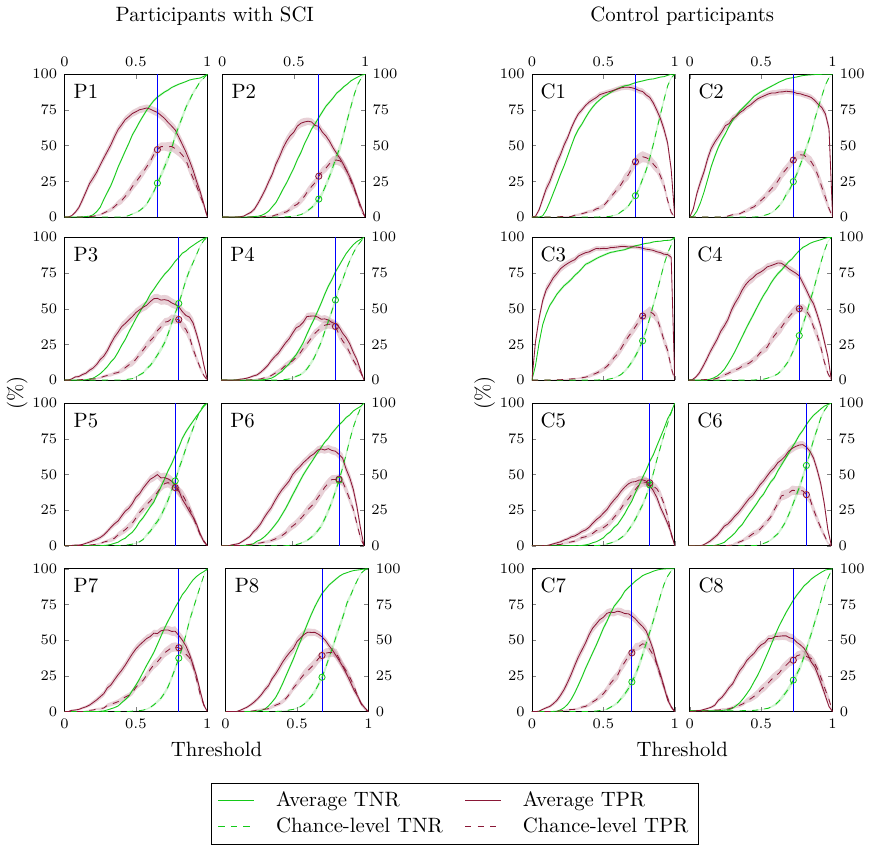}
	\caption{Evaluation of the personalized ErrP classifier. Single subject average TNR and TPR (green and red solid lines, respectively) and  chance level TNR and TPR in function of the threshold (green and red dashed lines, respectively).  The shaded areas indicate the \SI{95}{\percent} confidence interval for the average curves. The blue vertical line indicates the threshold that maximizes the individual performance.}
	\label{fig:cv_ss}
\end{figure}

\begin{figure}[h]
	\centering
		\includegraphics{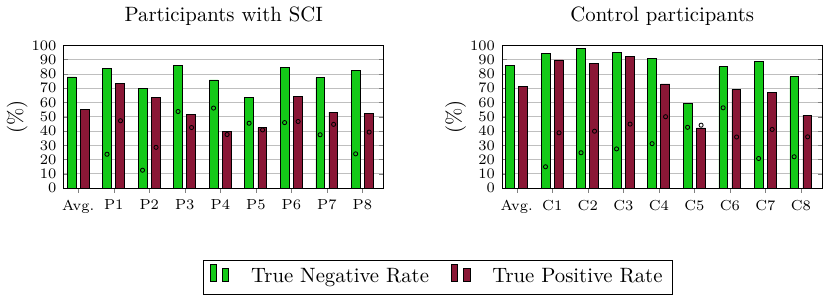}
	\caption{Evaluation of the personalized ErrP classifier. Average TNR and TPR calculated from the cross-validation procedure, with the optimal personalized threshold, for every participant and their average. The small dots on each bar indicate the chance level with the considered threshold, for every participant.}
	\label{fig:cv_ss_optimal}
\end{figure}

Figure~\ref{fig:cv_ss_optimal} depicts the average TNR and TPR  obtained in the cross-validation, when using the optimal personalized decision threshold for every participant  (green and red bars, respectively). The small circles on the bars indicate the chance level obtained for every participant with the considered threshold. For participants with SCI, the grand average TNR was \SI{77.9}{\percent} and the grand average TPR was \SI{55.0}{\percent}. For control participants, the grand average TNR was \SI{86.1}{\percent} and the grand average TPR was \SI{71.5}{\percent}.

\section{Discussion}

In this work we investigated the transfer of a generic ErrP classifier from able-bodied participants to participants with SCI.
The classifier was developed using the data from able-bodied participants from a previous experiment of ours \cite{Lopes-Dias2019} and was tested asynchronously in a closed-loop online experiment in which participants with SCI and able-bodied control participants took part. Using the classifier asynchronously, the entire trials were evaluated and not only a time-locked window. The online experiment required no offline calibration period and the participants received feedback of the ErrP detections immediately from the start of the experiment onwards. Additionally, we also analysed the morphology of error-related potentials in participants with SCI and in able-bodied control participants.

The grand average correct signal displayed, as expected, no particular potential both in participants with SCI and in control participants. The correct epochs correspond to the period in which the participants were continuously controlling the robot and were not associated with any specific event. 
The grand average error signal was associated with a fronto-central activity both in participants with SCI and control participants. The peaks of the grand average error signal were less pronounced in participants with SCI than in control participants, as visible in Figure~\ref{fig:ga}. This matches the results described in \cite{keyl}. Nevertheless, the electrophysiological patterns of participants with SCI were rather heterogeneous and half of the participants with SCI did not display the characteristic error-related activity (participants P4, P5, P7 and P8). The remaining participants with SCI revealed patterns comparable to control participants. Therefore, we believe that in our study the decrease in peak amplitudes observed in the grand average error signal of participants with SCI is not directly related with the injury, but rather a consequence of heterogeneity of the signals in the population with SCI. Several studies reported the effect of psychological factors, such as depression and anxiety, on error-related potentials \cite{RUCHSOW200637,OLVET201030}. The population with SCI is particularly vulnerable to emotional disorders and higher levels of distress \cite{Khazaeipour2015, migliorini2008, Post2012}. Nevertheless, individual differences are large \cite{Post2012}. To consider a psychiatric evaluation and medication of the participants would have been valuable for the current work and should be considered in future studies involving a population with SCI  \cite{vanLeeuwen2012}. Interestingly, the error signals of control participants were also less homogeneous than in our previous studies with a similar experimental protocol \cite{Lopes_Dias_2018, Lopes-Dias2019}. Several studies showed that ageing affects error processing and consequently the error-related potentials, hence we hypothesise that the higher variability observed in the signals of control participants of this study is related to the wider age range of the participants in comparison with previous studies \cite{Falkenstein2001,Hoffmann2011,Nieuwenhuis2002}.


  In order to interpret the classification results we focus on the TPR. This metric considers an interval after the error onset and the period before the error onset. Hence, it translates not only the classifier's ability to decode ErrPs after the occurrence of an error but also its ability to not detect ErrPs when no error occurs. The TNR only captures the  classifier's ability to not detect ErrPs when no error occurs. It is still a meaningful metric but the TNR's outcome can be artificially increased by the use of a high decision threshold, as depicted in Figure~4 of the supplementary material. The classification results of the generic classifier were, on average, lower in participants with SCI than in control participants. Only half of the participants with SCI obtained a TPR above chance level. These participants were the ones that displayed clear error patterns.  In the control participants, seven out of eight participants obtained a TPR above chance level. The remaining participant (participant C5) did not obtain a TPR above chance level and  did not display a very clear error signal. Summarizing, all participants that displayed clear ErrP patterns in the electrophysiological analysis obtained above chance level results with the generic classifier, independently of the group (SCI or control). It would be rather interesting to further investigate the factors that affect the error patterns, independently of the spinal cord injury. These results support that  using a generic ErrP classifier is a valuable option to give immediate feedback to the participants. Moreover, it indicates that ErrPs are transferable across participants, and that the transfer can be applied to distinct populations, such as participants with SCI.

 With the generic classifier developed, participants received real-time feedback of the ErrP detections from the beginning of the experiment. Still, the first three blocks of the experiment were used to update the threshold applied to the generic classifier. We made this choice because we had previously shown that some participants strongly benefit from combining the generic ErrP classifier with a personalized decision threshold \cite{LopesDias_generic2019}. For most participants, the threshold was relatively stable after the first block. This supports the use of a personalized threshold with the generic classifier, as suggested in \cite{LopesDias_generic2019}. 
 	In a real-world online application, the occurrence of errors can not be easily assessed, since it is determined by a subjective perception of the BCI user. Such a constraint hinders an objective evaluation of any ErrP classifier, unless the participants can use a motor-based strategy to report the occurrence of errors. Still, our approach could be applied to a real-world asynchronous situation in which the occurrence of errors is unknown.  Nevertheless, in order to establish a personalized decision threshold, our approach would need, beforehand, a short online application in which the occurrence of errors is known. Such application could be the equivalent of one of our experimental blocks, which contained 9 errors and lasted less than 5 minutes.
 	
 In our experiment, we only gave participants feedback of the ErrPs detected after the error onset. This aimed to assure that participants experienced the same number of errors and had comparable expectations regarding the occurrence of errors. Providing participants with feedback of the false positive ErrP detections would have brought our experiment closer to a real-world application at the cost of putting participants in dissimilar circumstances, given that false positive ErrP detections could affect their behaviour and the generation of ErrPs.  For instance, participants with many false positive ErrP detections would certainly be affected by the feedback in a negative manner. Either by losing engagement or by disregarding the feedback. Such participants would no longer perceive the errors as meaningful and relevant and this could alter their ErrPs.

When testing offline, the asynchronous classification with a personalized classifier, two participants with SCI (participants P4 and P5) and one control participant (participant C5) obtained chance level TPR results. This indicates that the signals of these participants were not sufficiently different to build a personalized classifier and these participants obtained chance level results with both generic and personalized classifiers.

The classification results obtained with the personalized classifier are not directly comparable with the results obtained with the generic classifier because the classifiers were evaluated on different datasets. In a real-world scenario, we could provide participants immediate feedback of their brain signals using a generic classifier, while collecting data to train a personalized classifier. Simultaneously, we could compare, at regular intervals, the performance of the personalized and generic classifiers and swap the generic classifier for a personalized classifier, once the latter would grant significantly better performance.

\section{Conclusion}
Our work shows that a generic ErrP classifier can be used, asynchronously and online, by participants with SCI and able-bodied participants. Moreover, the generic ErrP classifier is transferable from an able-bodied population to a population with SCI. The developed classifier required no previous calibration with the participant and granted immediate feedback of the ErrP detections. Therefore, our findings can help to widespread the incorporation of ErrPs in BCIs for different types of users.


\subsection{Acknowledgements}
This work was supported by Horizon 2020 ERC Consolidator Grant 681231 ‘Feel Your Reach’. The authors would like to thank Reinmar J. Kobler, Lea Hehenberger and Joana Pereira for the fruitful discussions.  The authors would also like to thank the staff and patients of the AUVA-Rehabilitationsklinik Tobelbad for the warm welcome.

\section{References}
\bibliographystyle{ieeetr}
\bibliography{mybib}{}

\begin{thebibliography}{10}

\bibitem{Brunner_horizon2020}
C.~Brunner, N.~Birbaumer, B.~Blankertz, C.~Guger, A.~K{\"u}bler, D.~Mattia,
  J.~del R~Mill\'{a}n, F.~Miralles, A.~Nijholt, E.~Opisso, N.~Ramsey,
  P.~Salomon, and G.~R. M{\"u}ller-Putz, ``{BNCI} {Horizon} 2020: towards a
  roadmap for the {BCI} community,'' {\em Brain-Computer Interfaces}, vol.~2,
  no.~1, pp.~1--10, 2015.

\bibitem{millan2010}
J.~d.~R. Mill{\'a}n, R.~Rupp, G.~M{\"u}ller-Putz, R.~Murray-Smith,
  C.~Giugliemma, M.~Tangermann, C.~Vidaurre, F.~Cincotti, A.~K{\"u}bler,
  R.~Leeb, C.~Neuper, K.~M{\"u}ller, and D.~Mattia, ``Combining
  brain–computer interfaces and assistive technologies: State-of-the-art and
  challenges,'' {\em Frontiers in Neuroscience}, vol.~4, p.~161, 2010.

\bibitem{WOLPAW2002}
J.~R. Wolpaw, N.~Birbaumer, D.~J. McFarland, G.~Pfurtscheller, and T.~M.
  Vaughan, ``{Brain–computer} interfaces for communication and control,''
  {\em Clinical Neurophysiology}, vol.~113, no.~6, pp.~767 -- 791, 2002.

\bibitem{schalk2000}
G.~Schalk, J.~R. Wolpaw, D.~J. McFarland, and G.~Pfurtscheller, ``{EEG}-based
  communication: presence of an error potential,'' {\em Clinical
  Neurophysiology}, vol.~111, no.~12, pp.~2138 -- 2144, 2000.

\bibitem{ferrez2005}
P.~W. Ferrez and J.~d.~R. Mill\'{a}n, ``You are wrong!: Automatic detection of
  interaction errors from brain waves,'' in {\em Proceedings of the 19th
  International Joint Conference on Artificial Intelligence}, IJCAI'05,
  pp.~1413--1418, 2005.

\bibitem{yeung2004}
N.~Yeung, M.~Botvinick, and J.~D~Cohen, ``The neural basis of error detection:
  Conflict monitoring and the error-related negativity,'' {\em Psychological
  review}, vol.~111, pp.~931--59, 11 2004.

\bibitem{chavarriaga2014}
R.~Chavarriaga, A.~Sobolewski, and J.~d.~R. Mill\'{a}n, ``Errare machinale est:
  the use of error-related potentials in brain-machine interfaces,'' {\em
  Frontiers in Neuroscience}, vol.~8, p.~208, 2014.

\bibitem{llera2011use}
A.~Llera, M.~A. van Gerven, V.~G{\'o}mez, O.~Jensen, and H.~J. Kappen, ``On the
  use of interaction error potentials for adaptive brain computer interfaces,''
  {\em Neural Networks}, vol.~24, no.~10, pp.~1120--1127, 2011.

\bibitem{llera2012adaptive}
A.~Llera, V.~G{\'o}mez, and H.~J. Kappen, ``Adaptive classification on
  brain-computer interfaces using reinforcement signals,'' {\em Neural
  Computation}, vol.~24, no.~11, pp.~2900--2923, 2012.

\bibitem{yousefi2019}
R.~Yousefi, A.~R. Sereshkeh, and T.~Chau, ``Online detection of error-related
  potentials in multi-class cognitive task-based bcis,'' {\em Brain-Computer
  Interfaces}, vol.~6, no.~1-2, pp.~1--12, 2019.

\bibitem{artusi}
X.~{Artusi}, I.~K. {Niazi}, M.-F. {Lucas}, and D.~{Farina}, ``Accuracy of a
  {BCI} based on movement-related and error potentials,'' in {\em 2011 Annual
  International Conference of the IEEE Engineering in Medicine and Biology
  Society}, pp.~3688--3691, Aug 2011.

\bibitem{Zhang_2015}
H.~Zhang, R.~Chavarriaga, Z.~Khaliliardali, L.~Gheorghe, I.~Iturrate, and J.~d.
  R.~M. ~, ``{EEG}-based decoding of error-related brain activity in a
  real-world driving task,'' {\em Journal of Neural Engineering}, vol.~12,
  p.~066028, nov 2015.

\bibitem{spueler2015error}
M.~Sp{\"u}ler and C.~Niethammer, ``Error-related potentials during continuous
  feedback: using {EEG} to detect errors of different type and severity,'' {\em
  Frontiers in Human Neuroscience}, vol.~9, p.~155, 2015.

\bibitem{kreilinger2016single}
A.~Kreilinger, H.~Hiebel, and G.~R. M{\"u}ller-Putz, ``Single versus multiple
  events error potential detection in a {BCI}-controlled car game with
  continuous and discrete feedback,'' {\em IEEE Transactions on Biomedical
  Engineering}, vol.~63, no.~3, pp.~519--529, 2016.

\bibitem{hakim}
H.~{Si-Mohammed}, C.~{Lopes-Dias}, M.~{Duarte}, F.~{Argelaguet}, C.~{Jeunet},
  G.~{Casiez}, G.~R. {Müller-Putz}, A.~{Lécuyer}, and R.~{Scherer},
  ``Detecting system errors in virtual reality using eeg through error-related
  potentials,'' in {\em 2020 IEEE Conference on Virtual Reality and 3D User
  Interfaces (VR)}, pp.~653--661, 2020.

\bibitem{parra2003}
L.~C. Parra, C.~D. Spence, A.~D. Gerson, and P.~Sajda, ``Response error
  correction - a demonstration of improved human-machine performance using
  real-time {EEG} monitoring,'' {\em IEEE Transactions on Neural Systems and
  Rehabilitation Engineering}, vol.~11, pp.~173--177, June 2003.

\bibitem{kreilinger2012error}
A.~Kreilinger, C.~Neuper, and G.~R. M{\"u}ller-Putz, ``Error potential
  detection during continuous movement of an artificial arm controlled by
  brain--computer interface,'' {\em Medical \& Biological Engineering \&
  Computing}, vol.~50, no.~3, pp.~223--230, 2012.

\bibitem{zander2016neuroadaptive}
T.~Zander, L.~Krol, N.~Birbaumer, and K.~Gramann, ``Neuroadaptive technology
  enables implicit cursor control based on medial prefrontal cortex activity,''
  in {\em Proceedings of the National Academy of Sciences of the United States
  of America}, vol.~113(12), pp.~14898--14903, Dez 2016.

\bibitem{SalazarGomez2017CorrectingRM}
A.~F. Salazar-Gomez, J.~DelPreto, S.~Gil, F.~H. Guenther, and D.~Rus,
  ``Correcting robot mistakes in real time using {EEG} signals,'' {\em 2017
  IEEE International Conference on Robotics and Automation (ICRA)},
  pp.~6570--6577, May 2017.

\bibitem{SuKyoung2017}
S.-K. Kim, E.~A. Kirchner, A.~Stefes, and F.~Kirchner, ``Intrinsic interactive
  reinforcement learning - using error-related potentials for real world
  human-robot interaction,'' {\em Scientific Reports (Sci Rep)}, vol.~7: 17562,
  12 2017.

\bibitem{mousavi2017}
M.~Mousavi, A.~S. Koerner, Q.~Zhang, E.~Noh, and V.~R. de~Sa, ``Improving motor
  imagery {BCI} with user response to feedback,'' {\em Brain-Computer
  Interfaces}, vol.~4, no.~1-2, pp.~74--86, 2017.

\bibitem{ehrlich2018}
S.~K. Ehrlich and G.~Cheng, ``Human-agent co-adaptation using error-related
  potentials,'' {\em Journal of Neural Engineering}, vol.~15, p.~066014, sep
  2018.

\bibitem{Spueler2012}
M.~Spüler, M.~Bensch, S.~Kleih, W.~Rosenstiel, M.~Bogdan, and A.~Kübler,
  ``Online use of error-related potentials in healthy users and people with
  severe motor impairment increases performance of a {P300}-{BCI},'' {\em
  Clinical Neurophysiology}, vol.~123, no.~7, pp.~1328 -- 1337, 2012.

\bibitem{Kobler2018}
R.~J. Kobler, A.~I. Sburlea, and G.~R. M{\"u}ller-Putz, ``Tuning
  characteristics of low-frequency eeg to positions and velocities in
  visuomotor and oculomotor tracking tasks,'' {\em Scientific Reports}, vol.~8,
  no.~1, p.~17713, 2018.

\bibitem{Ofner2012}
P.~{Ofner} and G.~R. {Müller-Putz}, ``Decoding of velocities and positions of
  {3D} arm movement from {EEG},'' in {\em 2012 Annual International Conference
  of the IEEE Engineering in Medicine and Biology Society}, pp.~6406--6409, Aug
  2012.

\bibitem{Inoue2018}
Y.~Inoue, H.~Mao, S.~B. Suway, J.~Orellana, and A.~B. Schwartz, ``Decoding arm
  speed during reaching,'' {\em Nature Communications}, vol.~9, no.~1, p.~5243,
  2018.

\bibitem{Nooh2011}
A.~A. Nooh, J.~Yunus, and S.~M. Daud, ``A review of asynchronous
  electroencephalogram-based brain computer interface systems,'' in {\em
  International Conference on Biomedical Engineering and Technology}, 2011.

\bibitem{Allison2012}
B.~Z. Allison, C.~Brunner, C.~Altstätter, I.~C. Wagner, S.~Grissmann, and
  C.~Neuper, ``A hybrid {ERD/SSVEP BCI} for continuous simultaneous two
  dimensional cursor control,'' {\em Journal of Neuroscience Methods},
  vol.~209, no.~2, pp.~299 -- 307, 2012.

\bibitem{omedes2013}
J.~Omedes, I.~Iturrate, and L.~Montesano, ``Detection of event-less error
  related potentials,'' in {\em Proceedings of IROS 2013 Workshop on
  Neuroscience and Robotics "Towards a robot-enabled, neuroscience-guided
  healthy society"}, 2013.

\bibitem{omedes2014asynchronous}
J.~Omedes, I.~Iturrate, and L.~Montesano, ``Asynchronous detection of error
  potentials,'' in {\em Proceedings of the 6th Brain-Computer Interface
  Conference 2014}, 2014.

\bibitem{omedes2015analysis}
J.~Omedes, I.~Iturrate, J.~Minguez, and L.~Montesano, ``Analysis and
  asynchronous detection of gradually unfolding errors during monitoring
  tasks,'' {\em Journal of Neural Engineering}, vol.~12, no.~5, p.~056001,
  2015.

\bibitem{omedes2015asynchronous}
J.~Omedes, I.~Iturrate, R.~Chavarriaga, and L.~Montesano, ``Asynchronous
  decoding of error potentials during the monitoring of a reaching task,'' in
  {\em 2015 IEEE International Conference on Systems, Man, and Cybernetics
  (SMC2015)}, pp.~3116--3121, Oct 2015.

\bibitem{LopesDias2017}
C.~Lopes-Dias, A.~I. Sburlea, and G.~R. M{\"u}ller-Putz, ``Error-related
  potentials with masked and unmasked onset during continuous control and
  feedback,'' in {\em 7th Graz Brain-Computer Interface Conference 2017},
  pp.~320--332, September 2017.

\bibitem{Lopes_Dias_2018}
C.~Lopes-Dias, A.~I. Sburlea, and G.~R. Müller-Putz, ``Masked and unmasked
  error-related potentials during continuous control and feedback,'' {\em
  Journal of Neural Engineering}, vol.~15, p.~036031, apr 2018.

\bibitem{Lopes-Dias2019}
C.~Lopes-Dias, A.~I. Sburlea, and G.~R. M{\"u}ller-Putz, ``Online asynchronous
  decoding of error-related potentials during the continuous control of a
  robot,'' {\em Scientific Reports}, vol.~9, no.~1, p.~17596, 2019.

\bibitem{iturrate2011}
I.~{Iturrate}, L.~{Montesano}, R.~{Chavarriaga}, J.~{del R. Millán}, and
  J.~{Minguez}, ``Minimizing calibration time using inter-subject information
  of single-trial recognition of error potentials in brain-computer
  interfaces,'' in {\em 2011 Annual International Conference of the IEEE
  Engineering in Medicine and Biology Society}, pp.~6369--6372, 2011.

\bibitem{iturrate2012}
I.~{Iturrate}, R.~{Chavarriaga}, L.~{Montesano}, J.~{Minguez}, and
  J.~d.~R.~{Millán}, ``Latency correction of error potentials between
  different experiments reduces calibration time for single-trial
  classification,'' in {\em 2012 Annual International Conference of the IEEE
  Engineering in Medicine and Biology Society}, pp.~3288--3291, Aug 2012.

\bibitem{Iturrate_2014}
I.~Iturrate, R.~Chavarriaga, L.~Montesano, J.~Minguez, and J.~Mill{\'{a}}n,
  ``Latency correction of event-related potentials between different
  experimental protocols,'' {\em Journal of Neural Engineering}, vol.~11,
  p.~036005, apr 2014.

\bibitem{SuKyoung2016}
S.~K. Kim and E.~A. Kirchner, ``Handling few training data: Classifier transfer
  between different types of error-related potentials,'' {\em IEEE Transactions
  on Neural Systems and Rehabilitation Engineering}, vol.~24, pp.~320--332,
  March 2016.

\bibitem{SuKyoung2013}
S.~K. Kim and E.~A. Kirchner, ``Classifier transferability in the detection of
  error related potentials from observation to interaction,'' in {\em 2013 IEEE
  International Conference on Systems, Man, and Cybernetics}, pp.~3360--3365,
  Oct 2013.

\bibitem{Bhattacharyya}
S.~Bhattacharyya, A.~Konar, D.~N. Tibarewala, and M.~Hayashibe, ``A generic
  transferable {EEG} decoder for online detection of error potential in target
  selection,'' {\em Frontiers in Neuroscience}, vol.~11, p.~226, 2017.

\bibitem{ehrlich2018_1}
S.~K. Ehrlich and G.~Cheng, ``A feasibility study for validating robot actions
  using {EEG}-based error-related potentials,'' {\em International Journal of
  Social Robotics}, Nov 2018.

\bibitem{schonleitner2020}
F.~M. {Schönleitner}, L.~{Otter}, S.~K. {Ehrlich}, and G.~{Cheng},
  ``Calibration-free error-related potential decoding with adaptive
  subject-independent models: A comparative study,'' {\em IEEE Transactions on
  Medical Robotics and Bionics}, vol.~2, no.~3, pp.~399--409, 2020.

\bibitem{LopesDias_generic2019}
C.~{Lopes-Dias}, A.~I. {Sburlea}, and G.~R. {Müller-Putz}, ``Asynchronous
  detection of error-related potentials using a generic classifier,'' in {\em
  8th International Brain Computer Interface Conference 2019}, Sep 2019.

\bibitem{LopesDias_generic2020}
C.~{Lopes-Dias}, A.~I. {Sburlea}, and G.~R. {Müller-Putz}, ``Generic
  error-related potential classifier offers a comparable performance to a
  personalized classifier,'' in {\em EMBC}, 2020.

\bibitem{keyl}
P.~Keyl, M.~Schneiders, C.~Schuld, S.~Franz, M.~Hommelsen, N.~Weidner, and
  R.~Rupp, ``Differences in characteristics of error-related potentials between
  individuals with spinal cord injury and age- and sex-matched able-bodied
  controls,'' {\em Frontiers in Neurology}, vol.~9, p.~1192, 2019.

\bibitem{Kumar2019}
A.~Kumar, Q.~Fang, J.~Fu, E.~Pirogova, and X.~Gu, ``Error-related neural
  responses recorded by electroencephalography during post-stroke
  rehabilitation movements,'' {\em Frontiers in Neurorobotics}, vol.~13,
  p.~107, 2019.

\bibitem{Kobler2017_eye}
R.~J. Kobler, A.~I. Sburlea, and G.~R. M{\"u}ller-Putz, ``A comparison of
  ocular artifact removal methods for block design based electroencephalography
  experiments,'' in {\em Proceedings of the 7th Graz Brain-Computer Interface
  Conference 2017}, Sep 2017.

\bibitem{Kobler2020}
R.~J. Kobler, ``Corneo-retinal-dipole and eyelid-related eye artifacts can be
  corrected offline and online in electroencephalographic and
  magneticencephalographic signals,'' {\em NeuroImage}, 2020.

\bibitem{blankertz2011}
B.~Blankertz, S.~Lemm, M.~Treder, S.~Haufe, and K.-R. Müller, ``Single-trial
  analysis and classification of {ERP} components — a tutorial,'' {\em
  NeuroImage}, vol.~56, no.~2, pp.~814 -- 825, 2011.
\newblock Multivariate Decoding and Brain Reading.

\bibitem{giraldi2008}
G.~A. Giraldi, P.~S. Rodrigues, E.~C. Kitani, J.~R. Sato, and C.~E. Thomaz,
  ``Statistical learning approaches for discriminant features selection,'' {\em
  Journal of the Brazilian Computer Society}, vol.~14, pp.~7--22, Jun 2008.

\bibitem{Lopez-Larraz2014}
E.~L{\'o}pez-Larraz, L.~Montesano, {\'A}.~Gil-Agudo, and J.~Minguez,
  ``Continuous decoding of movement intention of upper limb self-initiated
  analytic movements from pre-movement eeg correlates,'' {\em Journal of
  NeuroEngineering and Rehabilitation}, vol.~11, p.~153, Nov 2014.

\bibitem{Sburlea_2015}
A.~I. Sburlea, L.~Montesano, and J.~Minguez, ``Continuous detection of the
  self-initiated walking pre-movement state from {EEG} correlates without
  session-to-session recalibration,'' {\em Journal of Neural Engineering},
  vol.~12, p.~036007, apr 2015.

\bibitem{pereira2018}
J.~Pereira, A.~I. Sburlea, and G.~R. M{\"u}ller-Putz, ``Eeg patterns of
  self-paced movement imaginations towards externally-cued and
  internally-selected targets,'' {\em Scientific Reports}, vol.~8, p.~13394,
  Sep 2018.

\bibitem{good2000}
P.~Good and P.~Good, {\em Permutation Tests: A Practical Guide to Resampling
  Methods for Testing Hypotheses}.
\newblock Springer series in statistics, Springer, 2000.

\bibitem{ojala2010}
M.~Ojala and G.~C. Garriga, ``Permutation tests for studying classifier
  performance,'' {\em J. Mach. Learn. Res.}, vol.~11, p.~1833–1863, Aug.
  2010.

\bibitem{fatourechi}
M.~Fatourechi, R.~Ward, and G.~Birch, ``Evaluating the performance of a
  self-paced bci with a new movement and using a more engaging environment,''
  {\em Conference proceedings: Annual International Conference of the IEEE
  Engineering in Medicine and Biology Society. IEEE Engineering in Medicine and
  Biology Society. Conference}, vol.~2008, pp.~650--3, 02 2008.

\bibitem{RUCHSOW200637}
M.~Ruchsow, B.~Herrnberger, P.~Beschoner, G.~Grön, M.~Spitzer, and M.~Kiefer,
  ``Error processing in major depressive disorder: Evidence from event-related
  potentials,'' {\em Journal of Psychiatric Research}, vol.~40, no.~1, pp.~37
  -- 46, 2006.

\bibitem{OLVET201030}
D.~M. Olvet, D.~N. Klein, and G.~Hajcak, ``Depression symptom severity and
  error-related brain activity,'' {\em Psychiatry Research}, vol.~179, no.~1,
  pp.~30 -- 37, 2010.

\bibitem{Khazaeipour2015}
Z.~Khazaeipour, S.-M. Taheri-Otaghsara, and M.~Naghdi, ``Depression following
  spinal cord injury: Its relationship to demographic and socioeconomic
  indicators,'' {\em Topics in spinal cord injury rehabilitation}, vol.~21,
  no.~2, pp.~149--155, 2015.
\newblock 26364284[pmid].

\bibitem{migliorini2008}
C.~Migliorini, B.~Tonge, and G.~Taleporos, ``Spinal cord injury and mental
  health,'' {\em Australian \& New Zealand Journal of Psychiatry}, vol.~42,
  no.~4, pp.~309--314, 2008.
\newblock PMID: 18330773.

\bibitem{Post2012}
M.~W.~M. Post and C.~M.~C. van Leeuwen, ``Psychosocial issues in spinal cord
  injury: a review,'' {\em Spinal Cord}, vol.~50, no.~5, pp.~382--389, 2012.

\bibitem{vanLeeuwen2012}
C.~M.~C. van Leeuwen, L.~H.~V. van~der Woude, and M.~W.~M. Post, ``Validity of
  the mental health subscale of the sf-36 in persons with spinal cord injury,''
  {\em Spinal Cord}, vol.~50, no.~9, pp.~707--710, 2012.

\bibitem{Falkenstein2001}
M.~Falkenstein, J.~Hoormann, and J.~Hohnsbein, ``Changes of error-related
  {ERPs} with age,'' {\em Experimental Brain Research}, vol.~138, no.~2,
  pp.~258--262, 2001.

\bibitem{Hoffmann2011}
S.~Hoffmann and M.~Falkenstein, ``Aging and error processing: age related
  increase in the variability of the error-negativity is not accompanied by
  increase in response variability,'' {\em PloS one}, vol.~6,
  pp.~e17482--e17482, Feb 2011.

\bibitem{Nieuwenhuis2002}
S.~Nieuwenhuis, K.~R. Ridderinkhof, D.~Talsma, M.~G.~H. Coles, C.~B. Holroyd,
  A.~Kok, and M.~W. van~der Molen, ``A computational account of altered error
  processing in older age: Dopamine and the error-related negativity,'' {\em
  Cognitive, Affective, \& Behavioral Neuroscience}, vol.~2, no.~1, pp.~19--36,
  2002.

\end{thebibliography}

\end{document}


\title{\small Supplementary material} 
\author{Catarina Lopes-Dias$^1$,  Andreea I. Sburlea$^1$,  Katharina Breitegger$^2$, Daniela Wyss$^2$, Harald Drescher$^2$,  Renate Wildburger$^2$ and Gernot R. Müller-Putz$^1$}


\vspace{1.0cm}

\noindent Figure~1 shows the location of the 61 EEG electrodes used in the experiment. Figure~2 displays the grand average of the features used to train the generic ErrP classifier. Figure~3 depicts the pattern of the generic ErrP classifier. Figure~4 displays the average TPR and TNR obtained from 500 generic classifiers in which the training labels were randomly permuted. Figure~5 defines the metric 'ErrP detection rate' and depicts it together with the true positive rate, for all participants. {\color{black}Table~1 presents the false activation rate of every participant for correct and error trials}. 
Figures~6 and 7 display the topographic plots for correct and error signals at different time points for participants with SCI and control participants.

\begin{figure}[H]
	\centering
	\includegraphics[width=0.6\linewidth]{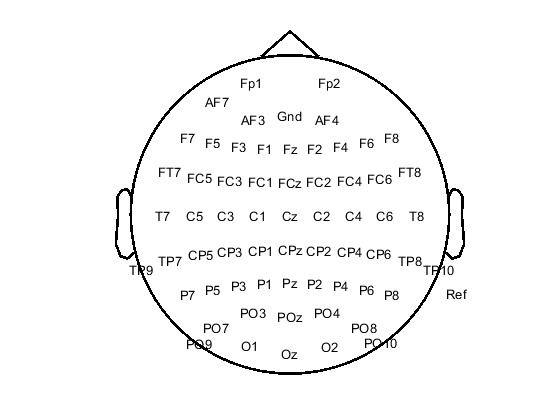}
	\caption{\textbf{Layout of the EEG electrodes:} Location of the 61 EEG electrodes used in the experiment. The ground electrode was placed at position AFz and the reference electrode was placed on the right mastoid.}
\end{figure}

\begin{figure}[H]
	\centering
	\includegraphics[width=0.9\linewidth]{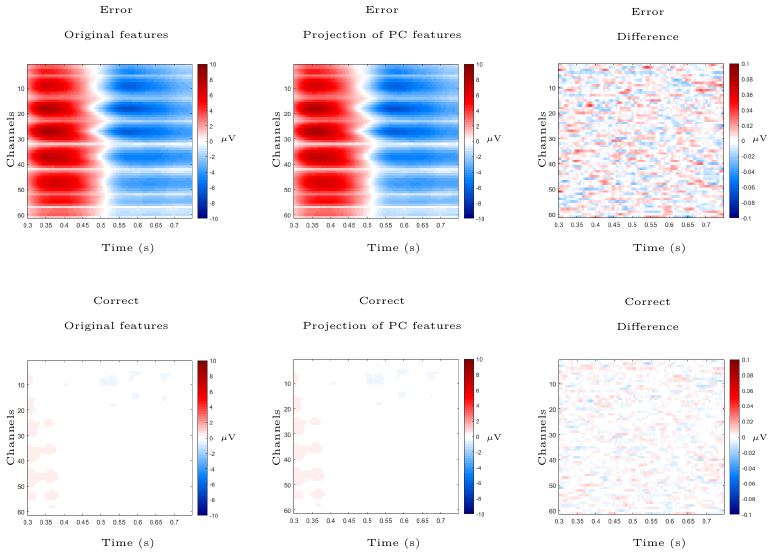}
	\caption{Grand average features of the generic classifier. \textbf{Left:} Grand average of the 1059 error epochs and of the 2475 correct epochs of the original feature space. \textbf{Middle:} Grand average of the projection into the temporal-spatial domain of the training matrix with 412 features, for error and correct epochs. The 412 features correspond to the principal components (PC) retained after PCA. \textbf{Right:} Difference
		between the grand average of the original feature space and the
		 grand average of the projection of PCA features. The channels are ordered from left to right and from front to back.}
\end{figure}

\begin{figure}[H]
	\centering
	\includegraphics[width=0.4\linewidth]{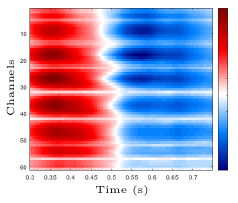}
	\caption{Generic classifier pattern: Pattern obtained by applying the discriminative feature analysis (DFA) method to the training matrix with 3534 trials and 412 features, used to train the shrinkage LDA classifier. The channels are ordered from left to right and from front to back.
	}
\end{figure}

\newpage

\begin{figure}[H]
	\includegraphics[width=\linewidth]{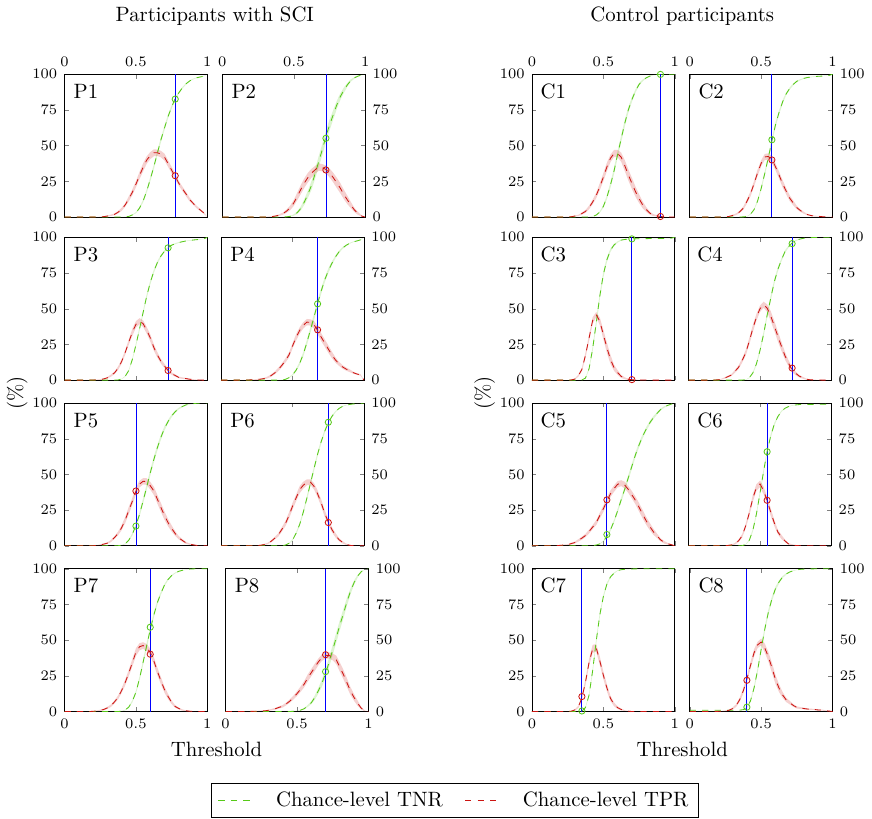}
	\caption{
		 Chance level of the generic ErrP classifier: Average TPR and TNR obtained from 500 generic classifiers in which the training labels were randomly permuted (dashed red and green lines, respectively). These classifiers were evaluated on blocks 4 to 8 of every participant. The shaded areas indicate the \SI{95}{\percent} confidence interval of the average curves. The vertical blue line indicates the threshold used for every participant during blocks 4 to 8 of the online experiment, obtained as described in Figure~5 of the manuscript. The chance levels depicted in Figure~6 of the manuscript, result from the intersection of the vertical threshold line with the average curves of TNR and TPR.}
\end{figure}

 \newpage

\begin{figure}[H]
	\vspace{1cm}
		\includegraphics[width=\linewidth]{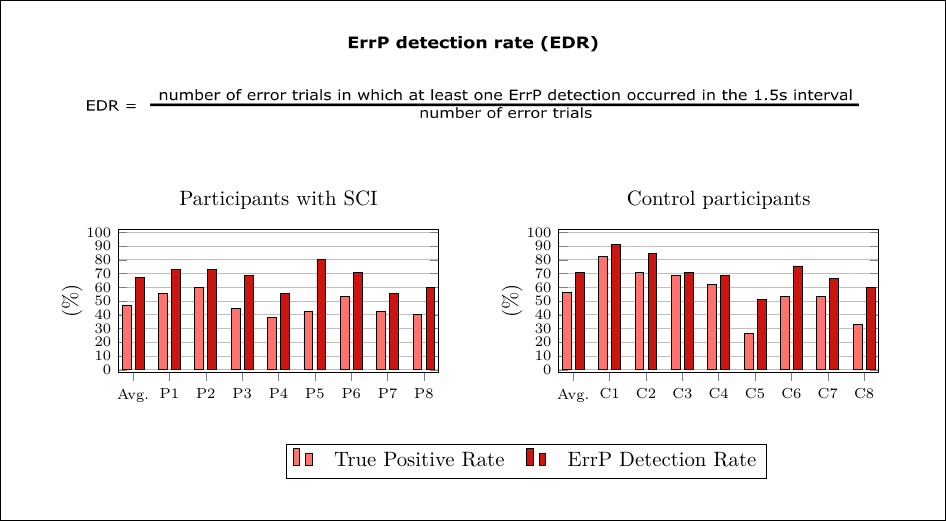}
	\caption{ErrP detection rate with the generic classifier: Definition of the metric ErrP detection rate (EDR) and its comparison with the metric true positive rate. The ErrP detection rate captures the portion of error trials in which at least one ErrP was successfully detected in the \SI{1.5}{\second} after the error onset, independently of the classifier output before the error onset.  The ErrP detection rate is less restrictive than the true positive rate. The difference between  ErrP detection rate and true positive rate captures the portion of error trials in which a false ErrP detection occurred before the error onset despite at least one successful ErrP detection also occurred in the assigned  \SI{1.5}{\second}. In participants with SCI, the average ErrP detection rate was \SI{67.2}{\percent}. In control participants, the average  ErrP detection rate was \SI{71.1}{\percent}.
	}
\end{figure}


\begin{table}[H]
	{\small
	\subfloat[\large Participants with SCI]{
		\begin{tabular}{c  c c c c c}
			\toprule
			\multicolumn{1}{c}{} &  \multicolumn{2}{c}{FAR (\%)} & & \multicolumn{2}{c}{Total nr of sec}  \\
			\cmidrule{2-3} \cmidrule{5-6} 
			&  {Correct} & {Error}& & {Correct} & {Error}\\ 
			\midrule
			P1 & 13.9 & 23.9 & &172 & 46 \\ 
			P2 & 12.1 & 14.0 & &215 & 50  \\ 
			P3 & 14.2 & 27.9 & &141 & 43  \\ 
			P4 & 20.4 & 25.0 & &167 & 40  \\ 
			P5 & 36.6 & 17.9 & &142 & 39  \\ 
			P6 & 11.0 & 23.7 & &118 & 38  \\ 
			P7 & 22.1 & 21.6 & &131 & 37  \\ 
			P8 & 16.8 & 32.5 & &154 & 40  \\ 
			\bottomrule
		\end{tabular} %
	}\hspace{0.3cm}
	\subfloat[\large Control Participants]{
		\begin{tabular}{c c  c c c c} 
			\toprule
			\multicolumn{1}{c}{} &  \multicolumn{2}{c}{FAR (\%)} & & \multicolumn{2}{c}{Total nr of sec}  \\
			\cmidrule{2-3} \cmidrule{5-6} 
			&  {Correct} & {Error} & & {Correct} & {Error}\\ 
			\midrule
			C1 & 3.5 & 9.1 & &173 & 44 \\ 
			C2 & 6.3 & 14.3 & &113 & 42  \\ 
			C3 & 2.3 & 4.7 & &174 & 43  \\ 
			C4 & 11.1 & 9.7 & &117 & 31  \\ 
			C5 & 27.9 & 33.3 & &147 & 39  \\ 
			C6 & 23.8 & 30.4 & &168 & 46  \\ 
			C7 & 25.2 & 20.9 & &147 & 43  \\ 
			C8 & 20.0 & 35.0 & &153 & 40  \\ 
			\bottomrule
	\end{tabular}}}\\
	\caption{{\color{black} False activation rate (FAR). The false activation rate in correct and error trials is denoted by FAR. The total number of 1-second long intervals evaluated in correct and error trials is denoted by Total nr of sec.}}
\end{table}

\newpage

\begin{figure}[h!]
	\includegraphics[width=\linewidth]{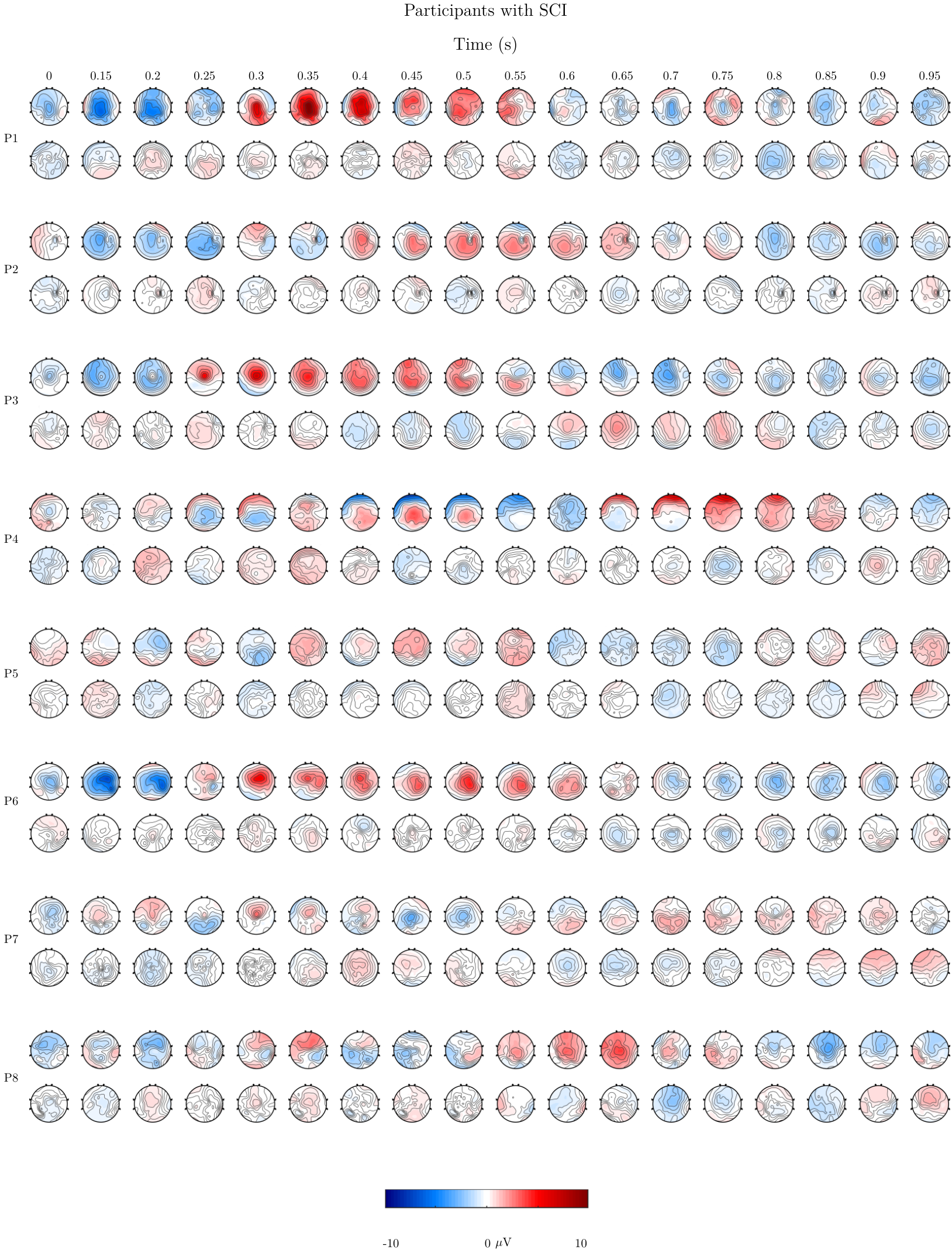}
	\caption{Participants with SCI: Topographic plots at different time points for error and correct signals (top and bottom row, respectively) for participants with SCI.}
\end{figure}

\begin{figure}[h!]
	\includegraphics[width=\linewidth]{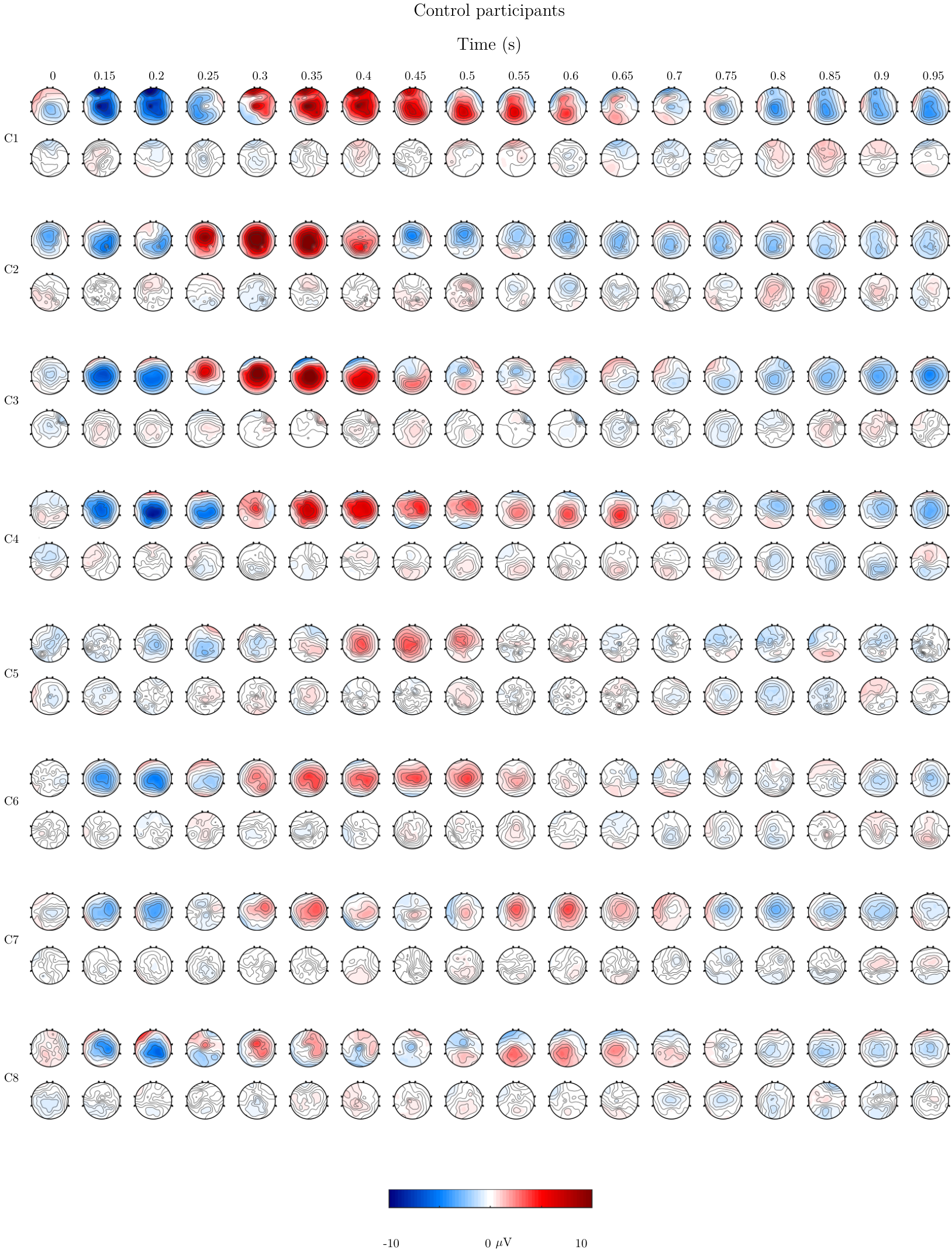}
	\caption{Control participants: Topographic plots at different time points for error and correct signals (top and bottom row, respectively) for control participants.}
\end{figure}